\documentclass[journal]{vgtc}                     

\onlineid{1028}

\vgtccategory{Leixian Shen, 
Leni Yang,
Haotian Li, 
Yun Wang, 
Yuyu Luo, 
and Huamin Qu}

\graphicspath{{figures/}{pictures/}{images/}{./}} 

\usepackage{colortbl}  
\usepackage{enumitem}
\setlist{noitemsep,parsep=0pt,partopsep=0pt, leftmargin=8pt} 
\usepackage{array} 
\usepackage{multirow} 
\usepackage{makecell} 
\usepackage{tabu}                      
\usepackage{booktabs}                  
\usepackage{lipsum}                    
\usepackage{mwe}                       

\usepackage{amsmath}
\usepackage{amssymb}
\usepackage{algorithm}
\usepackage[noend]{algpseudocode}
\usepackage{color}

\usepackage[x11names,HTML,svgnames]{xcolor}

\newcommand{\stitle}[1]{\noindent{\bf #1}}
\newcommand{\utitle}[1]{\noindent{\it \underline{#1}}}

\newcommand{\eg}{{\it e.g.,\ }}
\newcommand{\etal}{{\it et al.\ }}
\newcommand{\etc}{{\it etc.}}
\newcommand{\ie}{{\it i.e.,\ }}

\usepackage{hhline}
\definecolor{tablerowcolor}{rgb}{0.667,0.667,0.667 }
\definecolor{tablerowcolor2}{rgb}{0,0,0}

\definecolor{ckey}{rgb}{0.031, 0.031, 0.682}
\definecolor{cvalue}{rgb}{0.2, 0.439, 0.145}

\definecolor{mygreen}{RGB}{0,176,80}
\definecolor{mygreen2}{RGB}{232, 245, 233}
\definecolor{myorange}{RGB}{197, 80, 31}
\definecolor{myblue}{RGB}{0, 76, 193}

\definecolor{leniblue}{HTML}{004469}

\definecolor{visual}{HTML}{e8efd9}
\definecolor{motion}{HTML}{fde7d5}
\definecolor{narrative}{HTML}{e2dce9}
\definecolor{audio}{HTML}{d6ebf2}
\newcommand{\visual}[1]{\cellcolor{visual}{#1}}
\newcommand{\motion}[1]{\cellcolor{motion}{#1}}
\newcommand{\narrative}[1]{\cellcolor{narrative}{#1}}
\newcommand{\audio}[1]{\cellcolor{audio}{#1}}

\title{How Does Empirical Research Facilitate Creation Tool Design? \\A Data Video Perspective}

\author{
\authororcid{Leixian Shen}{0000-0003-1084-4912}, 
\authororcid{Leni Yang}{0000-0003-4527-4905},
\authororcid{Haotian Li}{0000-0001-9547-3449},
\authororcid{Yun Wang}{0000-0003-0468-4043},
\authororcid{Yuyu Luo}{0000-0001-9530-3327},
and  \authororcid{Huamin Qu}{0000-0002-3344-9694}
}

\authorfooter{

\item
L. Shen and H. Qu are with HKUST.
lshenaj@connect.ust.hk; huamin@cse.ust.hk
\item
L. Yang is with Inria.
leni.yang@inria.fr 
\item
H. Li and Y. Wang are with Microsoft Research Asia.
\{haotian.li, wangyun\}@microsoft.com 
\item
Y. Luo is with HKUST (Guangzhou). 
yuyuluo@hkust-gz.edu.cn
}

\abstract{
Empirical research in creative design deepens our theoretical understanding of design principles and perceptual effects, offering valuable guidance for innovating creation tools.
However, how these empirical insights currently influence the development of creation tools, and how their integration can be enhanced in the future, remains insufficiently understood.
In this paper, we aim to unveil the gap through a case study on \textit{data videos}, a prominent and wide-spread medium for effective data storytelling. 
To achieve the goal, we conducted a comprehensive analysis of 46 empirical research papers and 48 creation tool papers on data video, complemented by interviews with 11 experts. 
Building upon a systematic collection and structured characterization of empirical research by their methodologies (\eg corpus analysis, comparative evaluations) and component focus (\eg visuals, motions, narratives, audio), we conducted a context-aware citation analysis and revealed a taxonomy of recurring patterns in how empirical findings inform tool design across citation functions (\eg problem framing, technical reference).
Expert interviews further uncovered researchers’ practice patterns in applying empirical findings (\eg adaptation, synthesis, iteration, \etc) and identified key factors influencing applicability, such as contextual relevance, granularity matching, clarity, credibility, and feasibility.
Finally, we derive suggestions and discuss future opportunities to foster closer mutual engagement between empirical and tool research, aiming to reinforce the theoretical grounding of creation tools and enhance the practical impact of empirical research.

} 


\teaser{
\centering
\includegraphics[width=\linewidth]{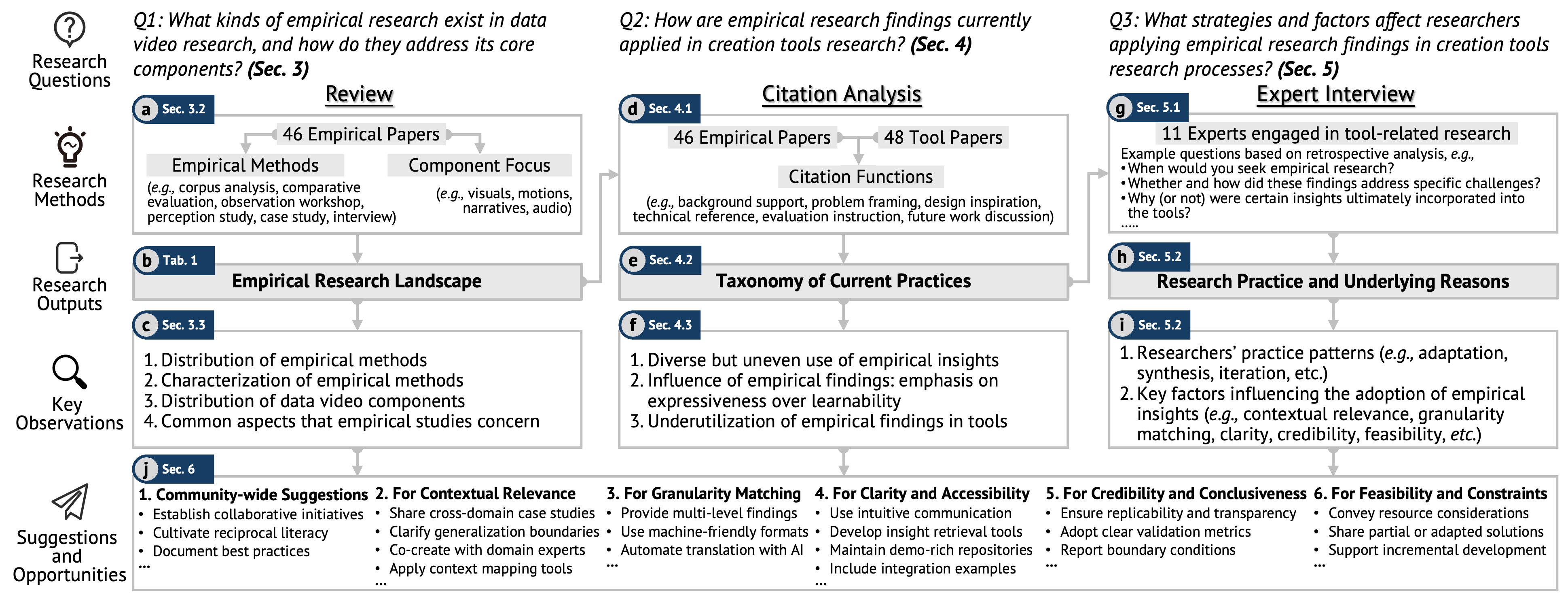}
\vspace{-10px}
\caption{
Overview of our research framework and logic, centered on three research questions, and organized by questions, methods, outputs, key observations, as well as suggestions and opportunities for future research and practice.
 }
\label{fig: teaser}
\vspace{-3px}
}

\begin{document}


\firstsection{Introduction}
\maketitle
Data videos have become a prevalent medium for data storytelling, integrating diverse multi-modal components to communicate insights effectively~\cite{DVSurvey, Amini2015}.
Due to their complexity and prevalence, significant research efforts have been put into promoting effective data videos.
There are two major research threads in academia.
One is \textit{empirical research}, which focuses on investigating best design practices~\cite{Dragicevic2011, Robertson2008}, perceptual effects~\cite{Kong2017a, Shu2020}, and storytelling strategies~\cite{Yang2022a, Lan2022} of data videos.
Another is to \textit{innovating interactive creation tool design} to facilitate the production of data videos efficiently and effectively, balancing expressiveness and learnability~\cite{DVSurvey}.
The two threads are not parallel but interact with each other in pursuit of the common goal (\ie prompting effective data videos). 
In one direction, designing creation tools applies empirical findings to enhance its theoretical foundations. 
In another direction, empirical research benefits from more study cases and richer study contexts enabled by advanced tools. 
They can provide twofold benefits: tool research gains evidence-based guidance to improve both user experience and data video quality, while empirical research achieves broad and impactful applications of its findings~\cite{Colusso2019}.

This paper focuses on enhancing our understanding and practices of the first direction, \ie applying empirical research findings in creation tools research, as it is essential and straightforward to complete the loop from theory to application.
Though there can be considerable benefits, it is also a challenging process. As noted by Colusso \etal\cite{Colusso2019}, translating empirical knowledge into research products often requires adapting abstract insights to technical constraints, reconciling design goals with research practices, and navigating cross-disciplinary collaboration.
For example, empirical research focuses on verifying specific phenomena or design strategies, which are often conducted in controlled conditions. It is foreseeable that creation tools research needs to adapt empirical recommendations to match their technical contexts and target audiences. 
However, we observed the lack of research in the current practices and influential factors in the application process, as well as future opportunities to improve it.
This lack of clarity points to a deeper need: not only to map what empirical research exists and how it is currently applied in creation tools research, but also to understand the practical strategies, factors, and barriers that shape its actual use.
The most relevant prior work, by Shen \etal~\cite{DVSurvey}, focuses on design paradigms for data video creation tools, but does not examine how empirical findings are applied in tool design.

To bridge this intra-academia gap, we take the initial step to answer the following research questions:
\begin{itemize}
    \item \textbf{Q1:} 
    What kinds of empirical research exist in data video research, and how do they address its core components?
    \item \textbf{Q2:} 
    How are empirical research findings currently applied in creation tools research?
    \item \textbf{Q3:} 
    What strategies and factors affect researchers applying empirical research findings in creation tools research processes?
\end{itemize}

By tackling Q1 and Q2, we sought to offer a quantitative overview of the knowledge landscape and current practices at the outcome level. 
In contrast, Q3 offers a qualitative perspective on implementation-level practices and the underlying reasons behind them.
Q3 complements Q1 and Q2 by revealing researchers’ iterative exploration and trial-and-error processes, while findings in Q1 and Q2 provided important references for answering Q3.
Taken together, these three questions form a holistic understanding of how empirical research can better inform tool design, guiding future progress on both sides.

To answer the three research questions, we conducted a mixed-method study, including literature review, citation analysis, and expert interviews, as shown in Fig.~\ref{fig: teaser}.
Specifically, to address Q1, we systematically reviewed the 46 empirical papers and organized them with two dimensions \textit{(a)}: the data video components studied, \ie visuals, motions, narratives, and audio, aligned with an existing creation tool reflection framework~\cite{DVSurvey}, and the empirical methods employed, including corpus analysis, comparative evaluation, observation workshop, perception study, case study, and interview~\cite{John2017}. 
This establishes an empirical research landscape \textit{(b)} and a set of observations of empirical research's characteristics \textit{(c)}, serving as a reference for future indexing and mapping to tool development needs.
Then, to answer Q2, we performed a citation analysis to examine how empirical findings are currently applied in creation tools research \textit{(d)}. We categorize citations into six functions: background support, problem framing, design inspiration, technical reference, evaluation instruction, and future work discussion~\cite{Hernandez2016}. By analyzing citation contexts across 48 tool papers and 46 empirical papers, we build a taxonomy of current practices across the citation functions \textit{(e)}. 
This taxonomy reveals a diverse but uneven pattern of empirical integration and suggests a disconnect between abundant available knowledge and practical underutilization \textit{(f)}.
Next, to explore Q3 and uncover the human-side reasons behind these patterns, we conducted in-depth interviews with 11 expert researchers who have built data video tool prototypes \textit{(g)}. We identified researchers' practice patterns when applying empirical findings (\eg adaptation, synthesis, iteration, \etc) and key factors (\eg contextual relevance, granularity matching, clarity, credibility, feasibility, \etc) that influence the adoption of empirical insights \textit{(h, i)}. 
Finally, combining all insights above, we derive suggestions and opportunities for the community \textit{(j)}.

Our main contributions are as follows:
\begin{itemize}

\item
We map the empirical research landscape of data videos by collecting and organizing studies according to methodology and component focus, serving as an index for future research.

\item 
We analyze citations between empirical and tool papers to build a taxonomy of how empirical insights are operationalized in creation tools research, identifying recurring patterns for future reference.

\item 
We conduct expert interviews to uncover strategies, factors, and barriers in applying empirical findings, and derive suggestions and opportunities for future research and practice.

\end{itemize}

\section{Related Work}

\subsection{Data Video}
The past 20 years have seen significant advancements in data videos.
Empirical research has explored various data video components, including visuals, motions, narratives, and audio, as well as their coordination, using diverse methods such as corpus analyses, interviews, and controlled experiments. 
For example, empirical research has derived and evaluated design options for visualization animations, transitions, and camera movements~\cite{Tang2020, Heer2007, Kim2019c, Shi2021b, Amini2018a, Yang2023a, DVEva}.
In addition, empirical research has examined different combinations of these components.
Examples include investigating how different components combine to achieve complex objectives, such as enhancing the delivery of narrative plots by audio-visual elements~\cite{Wei2024,Yang2022a,Lan2021a,Dasu2024}, achieving cinematic styles~\cite{Xu2022,Xu2023b,Conlen2023}, and eliciting emotional responses~\cite{Lan2022, Sakamoto2022}.

Over time, creation tools have also been proposed to address increasingly complex data video components. Early systems incorporated basic animation units with visualizations~\cite{Kim,Kim2020,Li2021c,Lu2020a}, whereas more recent efforts coordinate multiple visualizations~\cite{Amini2017,Lee,Shi2021a,DataParticles2023} or integrate real-world scenes~\cite{Chen2022c,Nam2024}. Developers have also extended these systems to include audio narration~\cite{Ying2023,dataplaywright, dataplayer,wonderflow} through text-to-speech features, sometimes leveraging music to improve engagement~\cite{Tang2022}. 

Although many data video creation tools have referenced empirical findings for conceptual grounding, the depth and nature of this integration varies widely, as some inform concrete features, while others serve only as background or are omitted entirely~\cite{DVSurvey}. 
Moreover, it remains unclear how researchers interpret, adapt, or selectively apply empirical research under real-world research and technical constraints.
This study addresses that gap by examining how empirical insights are applied in tool design, aiming to support their more effective use in driving practical innovation.

\subsection{Linking Empirical Insights to Tools}
The term ``intra-academia gap'' describes the disconnect of two closely related research areas, which hinders their co-evolution. Understanding and filling the gap is key to fostering the synergy between them~\cite{Colusso2019}.
This paper focuses on data video, a domain enriched with two lines of work: empirical research and creation tools research. 
As Shen \etal\cite{DVSurvey} noted, existing academic \emph{data video} tools seldom fully integrate empirical insights, given the computational and contextual demands of tool implementation and the fragmented nature of available findings.
As one of the most common data story genres, understanding the empirical-tool gap in data video research can both benefit the area and provide a probe into broader data visualization and storytelling research.

However, to our knowledge, no prior work has specifically investigated intra-academia gaps within the visualization community, providing limited references for our study.
Therefore, we seek inspiration from a related line that examines the disconnection between academic research and real-world application, \ie the research-practice gap.
These studies often employ strategies like literature review~\cite{Colusso2019, Velt2020}, citation analysis~\cite{Cao2023}, interview~\cite{Yildirim2023,Kumar2018}, \etc, to understand and address the gap with joint efforts.
Similarly, in the visualization community, researchers have also noted the difficulties in translating theoretical insights into functional systems, with designers often relying on ad-hoc or precedent-based knowledge over formal research~\cite{Parsons2021,Kim2023e,Wu}.

Inspired by these studies, we employ literature review, citation analysis, and expert interviews 
to quantitatively and qualitatively investigate: ``\textit{How does empirical research facilitate creation tool design?}'' 
Ultimately, our goal is to clarify how best to connect empirical insights and tool innovation, thereby making research findings more actionable and strengthening the academic ecosystem for data video research.

\subsection{Citation Analysis}
Citation analysis is a common bibliometric method that quantifies knowledge flow within and across research communities~\cite{Sajovic2022,Cao2023,Henry2007,Mannocci2019,Heimerl2016}. By identifying which studies are cited, and by whom, researchers can evaluate disciplinary impact and highlight the core ideas that shape subsequent work.
In HCI and visualization communities, citation analysis has been used to track domain evolution~\cite{Sajovic2022, Mannocci2019, Henry2007}, explore research impact~\cite{Cao2023}, and examine patterns of research adoption~\cite{Heimerl2016}. For example, Cao \etal\cite{Cao2023} investigated the impact of HCI research on industry based on patent citation analysis.

Given the large body of empirical and tool-based research papers on data videos, citation data can offer a lens into how empirical insights are adopted in creation tool development. It reveals which findings are reused, how they are applied, and where gaps persist. 
We follow and extend this line of work by analyzing how data video empirical papers are cited within creation tool research based on corresponding contexts~\cite{Hernandez2016}, offering a new perspective on knowledge transfer in HCI.

\section{Empirical Research Landscape}
\label{sec:knowledge}
In this section, we comprehensively collected papers related to data videos and systematically organized the empirical studies to answer \textit{Q1: What kinds of empirical research exist in data video research, and how do they address its core components?}

\subsection{Data Collection}
\label{sec: corpus}
\begin{figure}[t!]
\centering
\includegraphics[width=0.85\linewidth]{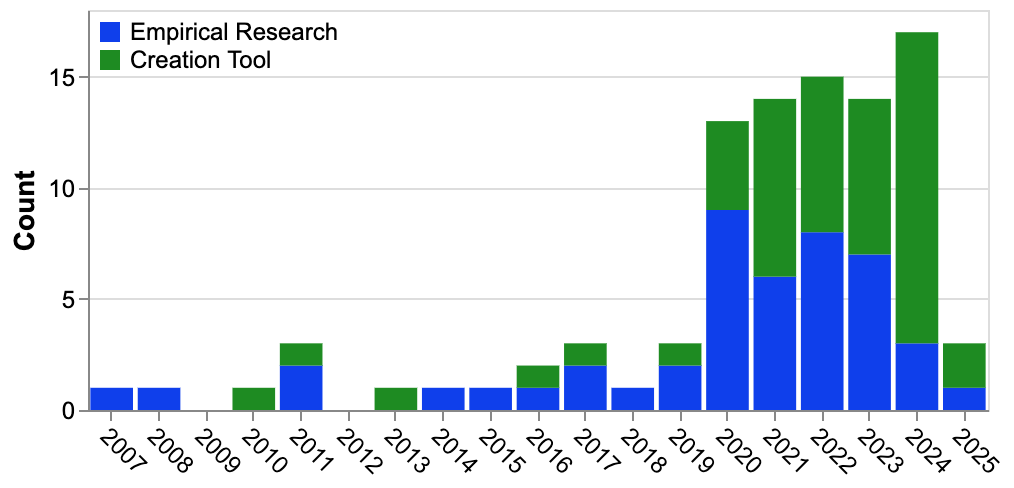}
\vspace{-5px}
\caption{Distribution of the 46 empirical research papers and 48 creation tool papers about data video in our corpus over time.
} 
\label{fig: time}
\vspace{-10px}
\end{figure}

We conducted keyword searches across multiple domains, including Visualization (VIS, EuroVis, Pvis, TVCG, CGF, CGA), Human-Computer Interaction (CHI, UIST, CSCW, IUI), and Computer Graphics and Multimedia (SIGGRAPH, MM), covering publications from 2000 to 2025. 
Following the definition of data videos~\cite{Amini2015}, we adopted similar search keywords and filtering criteria as in~\cite{DVSurvey} to establish an initial corpus, focusing on data-driven videos for storytelling presented on 2D screens.
To expand the corpus, we further reviewed references of related papers. 
Ultimately, we collected 46 empirical research papers and 48 creation tool papers. 
The distribution over time is shown in Fig.~\ref{fig: time}, showing increasing interest in data video research.

\subsection{Empirical Research Organization}\label{sec:empiricalOrg}
We first systematically organized the 46 empirical research papers to better understand their scope and relevance, applying two levels of categorization, as shown in Tab.~\ref{tab:Corpus}.  

First, we aligned our categorization with the framework proposed for understanding data video creation tools~\cite{DVSurvey}, organizing studies based on the data video components they address. By mapping empirical papers onto the same dimensions tool developers commonly work with, \ie visual (including visualization, pictograph, and real-world scene), motion (animation, transition, and camera), narrative (focusing on relationships between visualizations and their connections to real-world scenes), and audio (audio narration, music, and sound effect), we make it simpler to reference, understand, and apply empirical research findings in tool development.

Second, we categorized each study by its empirical methods, following guidance on classifying HCI research approaches~\cite{John2017}. Specifically, we examined each paper to identify distinct methods based on explicit descriptions of the protocol.
These research methods and their corresponding typical research output include:

\begin{itemize}
\item 
\textit{Corpus Analysis (C)}: Analyze collected cases or datasets to identify common patterns, often formulating structured design space, taxonomies, or guidelines~\cite{Wei2024, Yang2022a}.  
\item 
\textit{Comparative Evaluation (E)}: Compare multiple techniques, strategies, or workflows under controlled conditions to assess their effectiveness. Findings typically yield clear performance metrics, recommended best practices, or ranked design strategies~\cite{Heer2007,Rodrigues2024}.
\item 
\textit{Observation Workshop (O)}: Conduct structured workshops with targeted users to observe their behaviors, workflows, and phenomena in realistic scenarios. Results commonly include observational data, user requirements, behavioral patterns, or design opportunities~\cite{Amini2015, Thompson2020}.
\item 
\textit{Perception Study (P)}: Investigate specific human perceptual and cognitive factors affecting users’ interaction with data videos. Outputs usually consist of perceptual thresholds, cognitive constraints, design principles, or human-centered recommendations for optimal visual and narrative presentation~\cite{Rubab2023,Chalbi2020}.
\item 
\textit{Case Study (S)}: Conduct detailed qualitative examinations of specific instances or phenomena. Typical outcomes include illustrative examples, nuanced insights, practical lessons learned, or in-depth explorations of innovative techniques or design contexts~\cite{Conlen2023,Lin2023a}.

\item 
\textit{Interview (I)}: Gather qualitative insights through structured or semi-structured conversations, capturing subjective user experiences, motivations, needs, or challenges~\cite{Chevalier2016,Sallam2022}. 
\end{itemize}

As shown in Tab.~\ref{tab:Corpus} (\textit{Method}), a single paper can incorporate multiple methods (\eg corpus analysis and observation workshop for \cite{Wei2024}), each method was counted as a separate instance, leading to more \textit{empirical method instances} (74) than papers (46). This finer-grained view ensures that the full range of insights contributed by each empirical approach is accurately represented.
Furthermore, different methods (\eg comparing multiple animation strategies vs. observing real-world user behaviors) tend to yield distinct types of empirical findings (\eg design spaces, guidelines, best design practices).
When combining multiple methods, a single paper often yields a mix of insight types and granularities. 
Grouping studies by method thus provides a high-level lens to understand what kinds of insights are generated and where they might be derived or applied.

\begin{figure}[t!]
\centering
\includegraphics[width=0.95\linewidth]{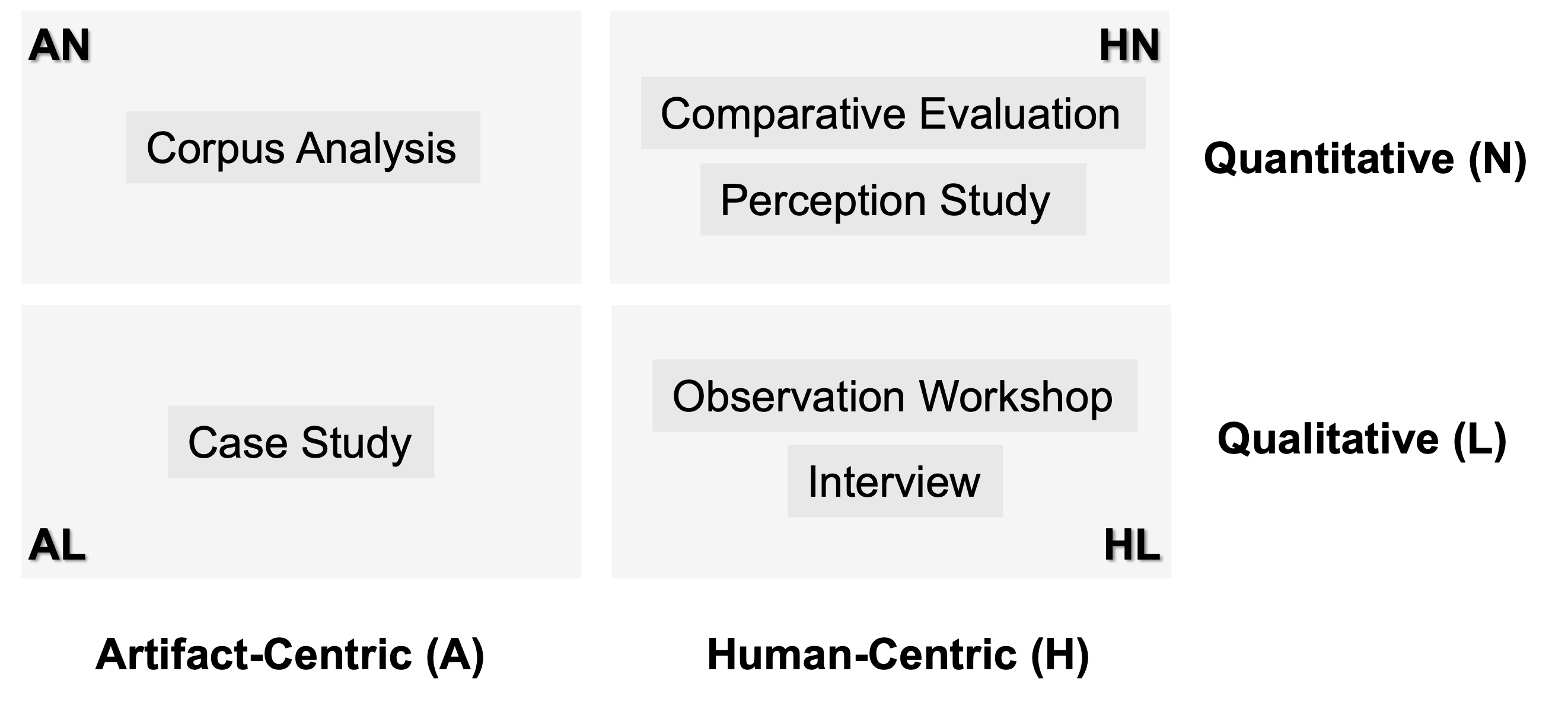}
\vspace{-10px}
\caption{Methods for deriving empirical findings, organized by two dimensions: qualitative/quantitative, human-centric/artifact-centric.
} 
\label{fig: method}
\vspace{-10px}
\end{figure}

\begin{table}[t!]
\small
\centering
\caption{Empirical research landscape. They are labeled with data video components summarized in~\cite{DVSurvey} and empirical methods, including Corpus Analysis (C), Comparative Evaluation (E), Observation Workshop (O), Perception Study (P), Case Study (S), and Interview (I).}
\label{tab:Corpus}
\setlength{\tabcolsep}{0.8mm}{
\renewcommand\arraystretch{1}
\begin{tabular}{ccc|ccc|ccc|c|ccc|}
\cline{4-13}
\multirow{2}{*}{} & 
\multirow{2}{*}{} & 
\multirow{2}{*}{} & 
\multicolumn{3}{c|}{\visual{\textbf{Visual}}} & 
\multicolumn{3}{c|}{\motion{\textbf{Motion}}} & 
\multirow{2}{*}{\narrative{\textbf{}}} & 
\multicolumn{3}{c|}{\audio{\textbf{Audio}}} \\ \cline{4-9} \cline{11-13} 
\textbf{Paper} & \textbf{Source} & \textbf{Method} & 
\rotatebox{90}{\visual{\textbf{Visualization}}} & 
\rotatebox{90}{\visual{\textbf{Pictograph}}} & 
\rotatebox{90}{\visual{\textbf{Real-World Scene}}}  & 
\rotatebox{90}{\motion{\textbf{Animation}}} & 
\rotatebox{90}{\motion{\textbf{Transition}}} & 
\rotatebox{90}{\motion{\textbf{Camera}}} & 
\rotatebox{90}{\narrative{\textbf{Narrative}}} & 
\rotatebox{90}{\audio{\textbf{Audio Narration}}} & 
\rotatebox{90}{\audio{\textbf{Music}}} & 
\rotatebox{90}{\audio{\textbf{Sound Effect}}}\\ 
\hline
Wei \etal\cite{Wei2024} & TVCG'25 & C/O & \visual{\checkmark} & \visual{\checkmark} &  & \motion{\checkmark} & \motion{\checkmark} & \motion{\checkmark} & \narrative{\checkmark} &  & \audio{\checkmark} & \audio{\checkmark} \\ \hline
Dasu  \etal\cite{Dasu2024} & TVCG'24 & C/S &  &  & \visual{\checkmark} &  &  &  & \narrative{\checkmark} &  &  &  \\ \hline
Rodrigues  \etal\cite{Rodrigues2024} & TVCG'24 & E & \visual{\checkmark} &  &  & \motion{\checkmark} & \motion{\checkmark} &  &  &  &  &  \\ \hline
Kim  \etal\cite{Kim2023b} & TVCG'24 & C & \visual{\checkmark} &  &  & \motion{\checkmark} &  &  &  &  &  &  \\ \hline
Conlen  \etal\cite{Conlen2023} & arXiv'23 & C/S & \visual{\checkmark} & \visual{\checkmark} & \visual{\checkmark} & \motion{\checkmark} &  & \motion{\checkmark} &  & \audio{\checkmark} & \audio{\checkmark} & \audio{\checkmark} \\ \hline
Mittenentzwei  \etal\cite{Mittenentzwei2023a} & CGF'23 & S/E &  & \visual{\checkmark} & \visual{\checkmark} &  &  &  & \narrative{\checkmark} &  &  &  \\ \hline
Rubab  \etal\cite{Rubab2023} & JoV'23 & P & \visual{\checkmark} &  &  &  &  &  &  &  &  & \audio{\checkmark} \\ \hline
Herath  \etal\cite{Herath2023} & MUM'23 & C/O & \visual{\checkmark} &  & \visual{\checkmark} & \motion{\checkmark} &  &  &  & \audio{\checkmark} &  & \audio{\checkmark} \\ \hline
Xu  \etal\cite{Xu2023b} & CHI'23 & C/O & \visual{\checkmark} & \visual{\checkmark} & \visual{\checkmark} & \motion{\checkmark} &  & \motion{\checkmark} &  & \audio{\checkmark} &  &  \\ \hline
Lin  \etal\cite{Lin2023a} & CGA'23 & S & \visual{\checkmark} &  & \visual{\checkmark} & \motion{\checkmark} &  &  &  &  &  &  \\ \hline
Yang  \etal\cite{Yang2023a} & Pvis'23 & C/O & \visual{\checkmark} & \visual{\checkmark} &  & \motion{\checkmark} & \motion{\checkmark} & \motion{\checkmark} &  &  &  &  \\ \hline
Yang  \etal\cite{Yang2022a} & TVCG'22 & C/O & \visual{\checkmark} & \visual{\checkmark} & \visual{\checkmark} & \motion{\checkmark} &  & \motion{\checkmark} & \narrative{\checkmark} &  &  &  \\ \hline
Xu  \etal\cite{Xu2022} & CHI'22 & C/O & \visual{\checkmark} & \visual{\checkmark} & \visual{\checkmark} & \motion{\checkmark} &  & \motion{\checkmark} &  & \audio{\checkmark} & \audio{\checkmark} & \audio{\checkmark} \\ \hline
Cheng  \etal\cite{Cheng2022} & CGF'22 & C & \visual{\checkmark} & \visual{\checkmark} &  & \motion{\checkmark} & \motion{\checkmark} & \motion{\checkmark} & \narrative{\checkmark} & \audio{\checkmark} &  &  \\ \hline
Lan  \etal\cite{Lan2022} & TVCG'22 & C/O & \visual{\checkmark} &  &  & \motion{\checkmark} &  & \motion{\checkmark} &  &  &  &  \\ \hline
VanDenBosch  \etal\cite{VanDenBosch2022} & IMX'22 & E/P &  & \visual{\checkmark} & \visual{\checkmark} &  &  &  & \narrative{\checkmark} &  &  &  \\ \hline
Sakamoto  \etal\cite{Sakamoto2022} & Pvis'22 & E & \visual{\checkmark} & \visual{\checkmark} &  & \motion{\checkmark} &  &  & \narrative{\checkmark} & \audio{\checkmark} &  &  \\ \hline
Sallam  \etal\cite{Sallam2022} & CHI'22 & P & \visual{\checkmark} & \visual{\checkmark} & \visual{\checkmark} & \motion{\checkmark} &  &  & \narrative{\checkmark} & \audio{\checkmark} & \audio{\checkmark} &  \\ \hline
Yao  \etal\cite{Yao2022} & TVCG'22 & C/P & \visual{\checkmark} &  & \visual{\checkmark} & \motion{\checkmark} &  &  &  &  &  &  \\ \hline
Fisher  \etal\cite{Fisher2021} & VIS'21 & E & \visual{\checkmark} &  &  & \motion{\checkmark} &  &  &  &  &  &  \\ \hline
Shi  \etal\cite{Shi2021b} & CHI'21 & C/O & \visual{\checkmark} & \visual{\checkmark} &  & \motion{\checkmark} &  &  & \narrative{\checkmark} &  &  &  \\ \hline
Dasu  \etal\cite{Dasu2021} & TVCG'21 & S/O & \visual{\checkmark} & \visual{\checkmark} & \visual{\checkmark} & \motion{\checkmark} &  &  & \narrative{\checkmark} &  &  &  \\ \hline
Crnovrsanin  \etal\cite{Crnovrsanin2021} & TVCG'21 & C/E & \visual{\checkmark} &  &  & \motion{\checkmark} & \motion{\checkmark} &  &  &  &  &  \\ \hline
Lan  \etal\cite{Lan2021a} & CHI'21 & I/C/O &  & \visual{\checkmark} & \visual{\checkmark} &  &  &  & \narrative{\checkmark} &  &  &  \\ \hline
Shu  \etal\cite{Shu2020} & TVCG'21 & C/I/P & \visual{\checkmark} & \visual{\checkmark} &  & \motion{\checkmark} &  &  &  &  &  &  \\ \hline
Brehmer  \etal\cite{Brehmer2020} & TVCG'20 & E & \visual{\checkmark} &  &  & \motion{\checkmark} &  &  &  &  &  &  \\ \hline
Concannon  \etal\cite{Concannon2020} & CHI'20 & O &  & \visual{\checkmark} & \visual{\checkmark} &  &  &  & \narrative{\checkmark} & \audio{\checkmark} &  &  \\ \hline
Chalbi  \etal\cite{Chalbi2020} & TVCG'20 & P & \visual{\checkmark} &  &  & \motion{\checkmark} & \motion{\checkmark} &  &  &  &  &  \\ \hline
Tang  \etal\cite{Tang2020a} & JoV'20 & O & \visual{\checkmark} &  & \visual{\checkmark} & \motion{\checkmark} &  &  &  &  &  &  \\ \hline
Bradbury  \etal\cite{Bradbury2020} & IV'20 & C/E &  & \visual{\checkmark} & \visual{\checkmark} &  &  &  & \narrative{\checkmark} & \audio{\checkmark} &  &  \\ \hline
Pereira  \etal\cite{Pereira2020} & VIS'20 & E & \visual{\checkmark} &  &  & \motion{\checkmark} & \motion{\checkmark} &  &  &  &  &  \\ \hline
Tang  \etal\cite{Tang2020} & VIS'20 & C/S & \visual{\checkmark} & \visual{\checkmark} & \visual{\checkmark} & \motion{\checkmark} & \motion{\checkmark} &  &  &  &  &  \\ \hline
Thompson  \etal\cite{Thompson2020} & CGF'20 & C/O & \visual{\checkmark} & \visual{\checkmark} &  & \motion{\checkmark} & \motion{\checkmark} & \motion{\checkmark} &  &  &  &  \\ \hline
Cao  \etal\cite{Cao2020a} & VI'20 & C/S & \visual{\checkmark} & \visual{\checkmark} & \visual{\checkmark} & \motion{\checkmark} & \motion{\checkmark} & \motion{\checkmark} & \narrative{\checkmark} & \audio{\checkmark} & \audio{\checkmark} &  \\ \hline
Kong  \etal\cite{Kong2019} & CHI'19 & P/I & \visual{\checkmark} &  &  & \motion{\checkmark} &  &  &  & \audio{\checkmark} &  &  \\ \hline
Kim  \etal\cite{Kim2019c} & CGF'19 & E & \visual{\checkmark} &  &  & \motion{\checkmark} & \motion{\checkmark} &  &  &  &  &  \\ \hline
Amini  \etal\cite{Amini2018a} & AVI'18 & E & \visual{\checkmark} &  &  & \motion{\checkmark} &  &  &  &  &  &  \\ \hline
Kong  \etal\cite{Kong2017a} & CGF'17 & I/C/P & \visual{\checkmark} &  &  & \motion{\checkmark} &  &  &  &  &  &  \\ \hline
Brehmer  \etal\cite{Brehmer2017} & TVCG'17 & C & \visual{\checkmark} & \visual{\checkmark} &  & \motion{\checkmark} & \motion{\checkmark} &  &  &  &  &  \\ \hline
Chevalier  \etal\cite{Chevalier2016} & AVI'16 & C/I & \visual{\checkmark} & \visual{\checkmark} &  & \motion{\checkmark} &  &  &  &  &  &  \\ \hline
Amini  \etal\cite{Amini2015} & CHI'15 & C/O & \visual{\checkmark} & \visual{\checkmark} & \visual{\checkmark} &  & \motion{\checkmark} & \motion{\checkmark} & \narrative{\checkmark} &  &  &  \\ \hline
Chevalier  \etal\cite{Chevalier2014} & TVCG'14 & E & \visual{\checkmark} &  &  &  & \motion{\checkmark} &  &  &  &  &  \\ \hline
Archambault  \etal\cite{Archambault2011} & TVCG'11 & E & \visual{\checkmark} &  &  & \motion{\checkmark} &  &  &  &  &  &  \\ \hline
Dragicevic  \etal\cite{Dragicevic2011} & CHI'11 & E & \visual{\checkmark} &  &  & \motion{\checkmark} & \motion{\checkmark} &  &  &  &  &  \\ \hline
Robertson  \etal\cite{Robertson2008} & TVCG'08 & E & \visual{\checkmark} &  &  & \motion{\checkmark} &  &  &  &  &  &  \\ \hline
Heer  \etal\cite{Heer2007} & TVCG'07 & E & \visual{\checkmark} &  &  & \motion{\checkmark} & \motion{\checkmark} &  &  &  &  &  \\ \hline
\hline
 \multicolumn{3}{c|}{Frequency}  & 40 &23  &19  & 37 &16  &11  &15  &11  &5  &5  \\ \hline

\end{tabular}}
\vspace{-10px}
\end{table}

\subsection{Observations}\label{sec:corpusOber}
\stitle{Distribution of Empirical Methods.}
For the six empirical methods, corpus analysis (24/74)\footnote{We use this format throughout the paper to indicate that 24 out of 74 instances fall into this category.} is most commonly used for understanding data videos, followed by comparative evaluation (16/74) and observation workshops(14/74), while perception studies (8/74), case studies (7/74), and interviews are less frequent (5/74). 
In addition, many empirical studies employ a combination of methods, and certain methodological combinations are frequently used, such as using corpus analysis to define a design space or guidelines, followed by observation workshops to examine participant behaviors and derive new insights~\cite{Yang2023a, Herath2023, Shi2021b}.

\stitle{Characterization of Empirical Methods.}
These empirical methods can be further classified along two key dimensions (Fig.~\ref{fig: method}), \ie human-centric~(H) vs. artifact-centric~(A), and qualitative~(L) vs. quantitative~(N), resulting in four distinct groups. Among them, artifact-centric + quantitative (AN, such as corpus analysis) typically uncovers large-scale patterns, whereas human-centric + qualitative (HL, \eg interviews and observation) tend to be exploratory and need-finding. Human-centric + quantitative (HN, \eg controlled perception experiments, comparative evaluations) emphasizes hypothesis-driven testing of user factors, and artifact-centric + qualitative (AL, like case studies) explores nuanced, context-specific scenarios in depth.

\stitle{Distribution of Data Video Components.}
We adopted the decomposition of data video components from the creation tool framework proposed by~\cite{DVSurvey}. As shown in the bottom of Tab.~\ref{tab:Corpus}, empirical studies decrease from visual and animation components to narrative and audio. This pattern aligns with the increasing complexity of features: visuals and animations are foundational, while narrative and audio demand higher-level coordination and integration. This also aligns with the state of tool development, as many tools support animated visualizations, but fewer incorporate audio capabilities and advanced narrative structure.

\stitle{Common Aspects that Empirical Studies Concern.}
Existing empirical research has examined diverse aspects of data video components, such as their definitions~\cite{Amini2015}, characteristics~\cite{Heer2007}, benefits~\cite{Amini2018a}, limitations~\cite{Kong2019}, inter-component interplay~\cite{Cheng2022}, design strategies~\cite{Xu2023b}, and production paradigms~\cite{Thompson2020}. 
However, most studies primarily focus on enhancing \textit{expressiveness}—identifying design choices that improve communicative effectiveness and proposing principles or guidelines for expressive video composition. 
In contrast, relatively few studies address how to reduce \textit{learnability} barriers, such as examining interaction designs or human-AI collaboration patterns.
Furthermore, we observe a scarcity of empirical research on a few elements, including real-world scenes, music, and sound effects. At the same time, as noted in~\cite{DVSurvey}, these elements are only supported by limited creation tools.
We suspect that this observation may suggest a mutually reinforcing gap: limited empirical exploration may constrain the design space for future tools, while the absence of supporting tools may, in turn, discourage empirical investigation into these underrepresented modalities.

\section{Citation Analysis}
\label{sec:citation}
The observations in the previous section suggest that empirical and tool-related research often exhibit similar patterns, implying their potentially close relationship.
Motivated by these observations, in this section, we performed a citation analysis of empirical and tool papers about data video to explore \textit{Q2: How are empirical research findings currently applied in creation tool research?}

\subsection{Data Labeling and Analysis}  
\label{sec:citationFunc}
\begin{figure}[t!]
\centering
\includegraphics[width=0.95\linewidth]{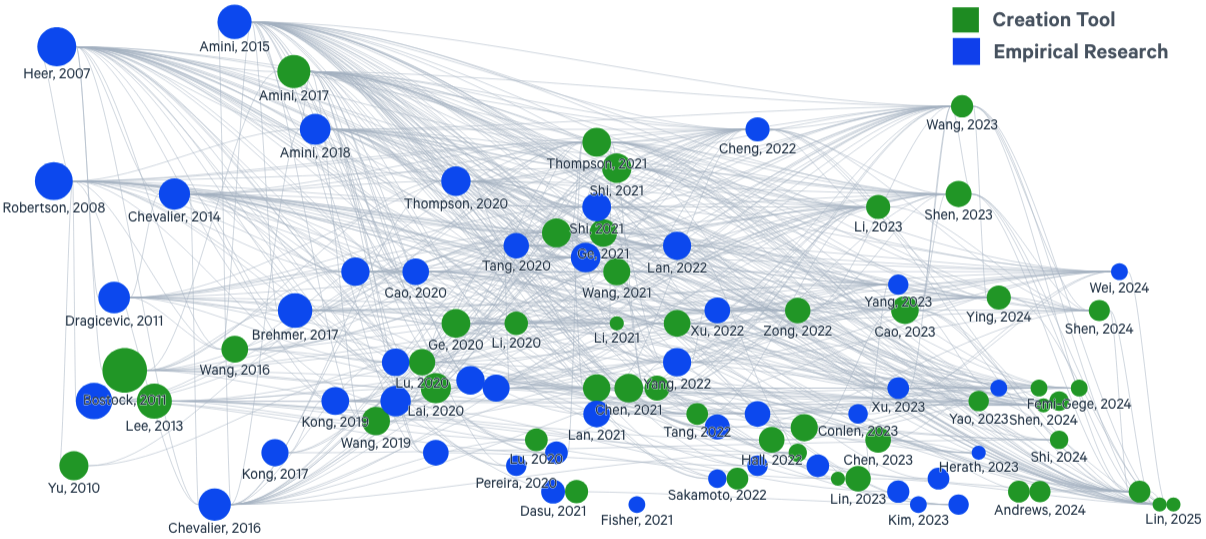}
\vspace{-5px}
\caption{
Citation graph of empirical and tool papers.
Node size indicates citation count; X-axis encodes year; Y-axis reflects topical relevance.
} 
\label{fig: citation}
\vspace{-10px}
\end{figure}

The citation graph of these papers is shown in Fig.~\ref{fig: citation}.
Inspired by existing literature about citation context analysis and scientific concept-based citation function classification~\cite{Hernandez2016}, we identify six citation functions commonly adopted in HCI about why creation tool papers reference empirical research: 
\begin{itemize}
\item 
\textit{Background Support}: Provide foundational knowledge or context.
\item 
\textit{Problem Framing}: Justify the research gap or motivate the development of creation tools.  
\item 
\textit{Design Inspiration}: Inform design considerations or requirements of new tools or interfaces.  
\item 
\textit{Technical Reference}: Guide technical solutions or implementation.
\item 
\textit{Evaluation Instruction}: Help establish evaluation methods or metrics.  
\item 
\textit{Future Work Discussion}: Highlight unresolved challenges or directions for future research.  
\end{itemize}

For each creation tool paper, we systematically examine all references to identify empirical research papers, locate all citation occurrences within the text, and analyze their surrounding contexts. 
Each citation is labeled with one of the six citation functions. 
During the process, one author initially annotated all data, while another author cross-checked the annotations, engaging in discussions to resolve discrepancies.
Since a single empirical research paper can serve multiple citation functions within a creation tool paper, it may receive multiple labels accordingly. 

\stitle{Distribution.}
Finally, we obtained 311 \textit{citation relationship instances} based on 46 empirical research papers and 48 creation tool papers.
Among them, \textit{background support} is the most frequently cited function (162), followed by \textit{design inspiration} (54). \textit{Problem framing}, \textit{technical reference}, and \textit{future work discussion} are cited less frequently, with 34, 32, and 22 instances, respectively. Evaluation instruction is the least frequently cited function, with only 7 instances. Detailed labeling data can be found in the supplementary materials.

Fig.~\ref{fig: heatmap} visualizes how often different types of empirical studies (\eg corpus analysis, comparative evaluation) are cited in creation tool papers for different purposes (\eg framing a research problem, informing design decisions). Each cell shows the average number of times a particular empirical method is cited under a specific function.
For example, ``perception study'' appears in 8 empirical papers, and these papers are collectively cited 4 times by tool papers specifically for ``problem framing.'' The corresponding cell shows an average of 0.5 citations (4 ÷ 8). This normalization helps us compare methods fairly, even if some are more common overall.
This metric helps reveal not just how often empirical research is cited, but how it contributes to tool development, highlighting the practical roles different research methods play.

\subsection{Taxonomy of Current Practice}
\label{sec:practice}
\begin{figure}[t]
\centering
\includegraphics[width=0.85\linewidth]{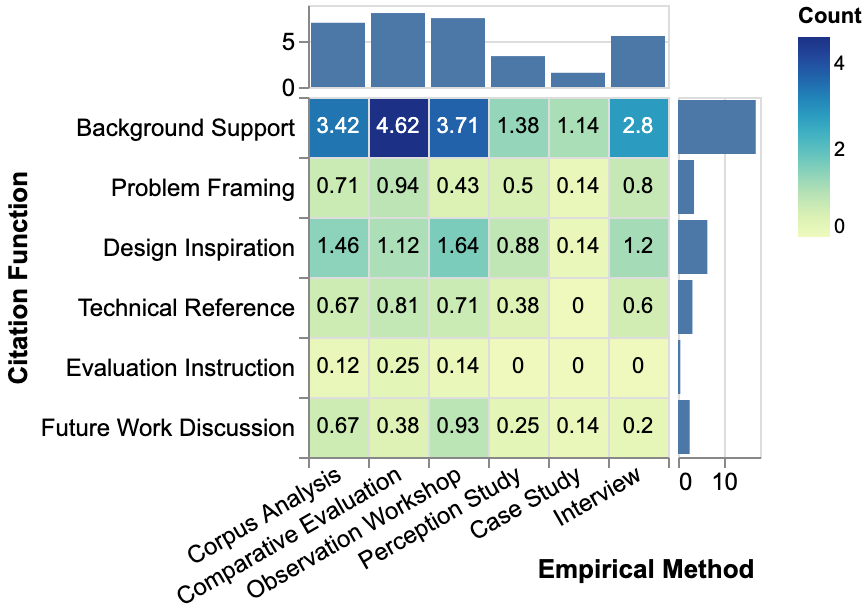}
\vspace{-5px}
\caption{Heatmap illustrating how frequently each empirical method is cited for each citation function on average across creation tool papers. 
} 
\label{fig: heatmap}
\vspace{-10px}
\end{figure}

We summarize common strategies and emerging patterns of how empirical findings were applied in creation tools under distinct citation functions, with a focus on problem framing, design inspiration, technical reference, and evaluation instruction, as they are more closely related to creation tool design. We also discuss the observed patterns in citation distribution (Fig.~\ref{fig: heatmap}).

\subsubsection{Background Support}
Background support (162/311) is the most prevalent citation function, where empirical research papers frequently serve to provide foundational knowledge and contextual information for creation tool papers.
They often explain core concepts (\eg animation transitions \cite{Heer2007,Chevalier2014}, narrative structures \cite{Yang2022a}) or summarize component benefits~\cite{Kong2017a,Brehmer2020}.
Although these citations enrich the reader’s understanding, most of them stop at problem awareness without prescribing a specific design route.

\textit{Reflection.} 
All empirical methods are widely applied for background support. 
Among all methods, 
comparative evaluations are especially prominent. 
The observation implies that researchers heavily leverage empirical research outcomes to support that their tool aims to address a real problem, indicating a tight coupling between empirical findings and problem definition.

\subsubsection{Problem Framing}
Problem framing (34/311) influenced the design of creation tools by motivating high-level research directions.

\stitle{Benefit-based Framing (25/34).}
In many cases, empirical studies emphasize the advantages of specific components, thereby motivating tool developers to integrate or extend these components.  
Quantitative methods (\eg corpus analysis, comparative evaluation, and perception studies) are particularly common here, as they systematically examine well-defined elements and verify their effectiveness.
For instance, the comparative effectiveness of animation transitions over static formats~\cite {Heer2007,Chevalier2014,Kim2019c,Archambault2011} motivated Gemini~\cite{Kim2020, Kim} and Canis~\cite{Ge2020} to build related animation grammars and recommendation mechanisms. 
Similarly, Cheng \etal\cite{Cheng2022} underscore the significance and patterns of narration-animation interplay by analyzing 426 clips in 60 data videos, inspiring the development of tools for audio-enriched data videos~\cite{wonderflow,dataplayer,dataplaywright}.
These findings function as ``green lights'' for system design, providing empirical proof that a component is worth supporting.

\stitle{Challenge-based Framing (9/34).}
Another group references prior research to demarcate challenges or identify user pain points.
Such citations often stem from qualitative sources (\eg interviews and observations) that illuminate user needs.
For example, Amini \etal\cite{Amini2015} identified challenges in composing diverse video components, motivating SmartShots~\cite{Tang2022} and AutoClips~\cite{Shi2021a}. 
Likewise, their observation on the dominance of prerecorded formats led to live-streaming explorations in DataTV~\cite{Zhao2022}. 
Here, empirical studies act as problem detectors, helping tools focus on concrete gaps.

\textit{Reflection.} 
Though less common than background-supporting citations, problem-framing references are key to aligning tool goals with empirically observed needs. Among these, human-centric methods (\eg interviews and comparative studies) are cited slightly more than artifact analyses, suggesting a preference for understanding user challenges. Still, the overall modest citation rate (averages mostly <1 across methods) implies that many authors do not ground their problem definitions in empirical research. 
While tools papers may justifiably build on prior systems or interaction design work, empirical studies can offer complementary benefits by surfacing recurring user challenges and overlooked opportunities specific to components. 
Expanding the scope of such studies, particularly towards rarely explored aspects like interactions and authoring pain points, can help better frame future tools.

\subsubsection{Design Inspiration}
Empirical studies often inspire creation tool designs by providing requirements and ideas (54/311).

\stitle{Feature Integration (10/54).}
A set of tools embeds key empirical insights as integral features, often drawing from comparative studies that systematically evaluate design alternatives. 
For example, 
DataClips~\cite{Amini2017} implements empirically validated authoring features, such as attention cues and data-driven templates, inspired by prior studies~\cite{Amini2015, Heer2007}. 
Likewise, 
Roslingifier~\cite{Shin2022} applies animation tracing techniques~\cite{Robertson2008} in its design space, and InfoMotion~\cite{Wang2021d} incorporates tested animation pacing strategies~\cite{Thompson2020}.
Such direct adoption ensures tools incorporate empirically validated best practices rather than ad hoc designs.

\stitle{Constraint Formalization (33/54).}
Empirical guidelines frequently serve as higher-level constraints or design considerations embedded within tool architectures.
Broadly validated principles, such as congruence and apprehension~\cite{Ferrand2010, Heer2007}, or specific animation transition guidance~\cite{Dragicevic2011, Chalbi2020, Chevalier2016}, inform fundamental design logics of animations across multiple tools~\cite{Shi2021a, Tang2022, Lee, Shen2024b, Pu2021, Wang2021d}. 
Tools like Animated Vega-Lite~\cite{Zong2022} also reflect static encoding practices~\cite{Shu2020} and handle more advanced abstraction layering~\cite{Thompson2020}.
SwimFlow \cite{Yao2024} adapts the video design guidelines of Tang \etal\cite{Tang2020a} from non-sports to the new sports contexts, demonstrating how generalized design principles can facilitate targeted innovation.
By formalizing these empirical insights, tool authors ensure their tools adhere to evidence-backed constraints, improving coherence and usability.

\stitle{Procedure Replication (9/54).}
A smaller but notable subset of tools replicates empirical study protocols or data processing pipelines.
For example, GeoCamera~\cite{Li2023b} replicated experimental protocols and datasets from prior empirical studies \cite{Tang2020, Shi2021b, Amini2015, Cheng2022} to develop their design space.  
Similarly, AutoClips~\cite{Shi2021a} references Amini \etal's\cite{Amini2015} data collection and analysis methods, while AniVis \cite{Li2021c} directly utilizes their dataset.  
Sporthesia \cite{Chen2022c} and VisCommentator~\cite{Chen2022h} follow dataset processing approaches from~\cite{Shu2020, Thompson2020, Amini2015}.  
Such replication promotes methodological rigor and ensures consistency with established empirical evidence.

\stitle{Scope Delimitation (2/54).}
A few tools leverage empirical findings to define what their tool will not do.
For example, Sporthesia~\cite{Chen2022c} excluded emotional cues in the design space due to the infancy of affective visualization research~\cite{Lan2022}.  
Hall \etal\cite{Hall2022} examine the strengths and limitations of animation transition~\cite{Heer2007} and restrict features to maintain synergy with a presenter’s body language.
This indicates that empirical work can shape boundaries as well as features.

\textit{Reflection.}
Empirical studies provide essential grounding for creation tool designs, facilitating evidence-supported innovations.
Tool authors draw inspiration from diverse empirical sources, leveraging both human-centric insights (user needs, experiences) and artifact-centric findings (established patterns, practices). Typically, quantitative studies frequently provide explicit ``do and don’t'' patterns, enabling straightforward integration as specific features or constraints. In contrast, qualitative studies offer rich, context-sensitive insights that inform higher-level system design considerations. 
However, uneven citation practices indicate that while some tools draw heavily on empirical evidence, many rely on other sources like prior tools or non-data-video studies, and empirical insights often remain underutilized.

\subsubsection{Technical Reference}
Empirical findings can also serve as technical references (32/311), guiding the implementation and architecture of creation tools. 

\stitle{Direct Implementation (11/32).}
Some tools directly implement technologies described in empirical research, usually grounded in comparative or perception studies that break down design components.
Animation grammars are a prime example, as they incorporate many fine-grained designs. 
For instance, Gemini~\cite{Kim2020} extends the staggering definition from Chevalier \etal\cite{Chevalier2014} and adopts a default slow-in/slow-out pacing strategy~\cite{Dragicevic2011}. 
Canis\cite{Ge2020} integrates Amini \etal’s\cite{Amini2015} animation unit facet and refines timing specifications using Chevalier~\etal’s~\cite{Chevalier2014} framework. 
Moreover, AutoClips~\cite{Shi2021a}  utilizes duration guidelines from Amini \etal~\cite{Amini2015} to optimize its duration configuration function, exemplifying empirical integration at the technical level.

\stitle{Taxonomy Adoption (10/32).}
Another pattern uses empirical taxonomies to guide tool logic.
Animated Vega-Lite~\cite{Zong2022} and DataClips~\cite{Amini2017}, for example, use animation transition taxonomies from Heer \etal\cite{Heer2007} to systematically analyze and implement design variations. 
Foreshadowing~\cite{Li2020c} and Vis-Annotator~\cite{Lai2020a} draw on visual cue taxonomies from Kong \etal\cite{Kong2017a}, while narration-based systems~\cite{NarrativePlayer,Ying2023} build on Cheng \etal’s\cite{Cheng2022} study on narration characteristics in data videos.
Thompson \etal’s\cite{Thompson2020} categorization of animation paradigms and their characteristics inspire keyframe-based~\cite{Pu2021, Wang2021d, Kim2020}, template-based~\cite{wonderflow}, and procedural-based systems~\cite{Lu2020a}.
These show how structured empirical insights can streamline complex design decisions.

\stitle{Knowledge Synthesis (11/32)}
A final approach synthesizes multiple empirical studies to form comprehensive design knowledge.
For example, Data Player~\cite{dataplayer} derives design constraints from prior studies~\cite{Shi2021b, Cheng2022, Amini2018a, Heer2007} to guide its constraint-based automatic animation sequence recommendation.  
GeoCamera~\cite{Li2023b} expands on narrative intent surveys from empirical research~\cite{Chevalier2016, Thompson2020, Shu2020}, integrating them into its interface for user selection.
Sporthesia~\cite{Chen2022c} and VisCommentator~\cite{Chen2022h} leverage prior studies~\cite{Shu2020, Thompson2020, Amini2015} to shape design spaces for advanced data-video augmentation.

\textit{Reflection.}
Empirical studies clearly inform technical implementations, though less extensively compared to motivational or inspirational citations. 
It shows a notable reliance on artifact-centric or quantitative methods, perhaps because these studies provide clear, modular findings (\eg grammar definitions, recommended timings) that map directly onto tool features. 
However, this focus also suggests that qualitative insights, which are often rich in context but less structured, may be overlooked due to their less computational applicability. 
Moreover, due to the inherent emphasis on innovation, creation tools rarely implement empirical findings as-is for their core features. Instead, these findings often support auxiliary functions or serve as foundational elements within larger design innovations. 
In addition, there is a natural progression as novel tools usually push technological boundaries where direct empirical precedents may not yet exist. Thus, from the technical perspective, empirical research functions more as a foundational reference rather than a prescriptive blueprint.

\subsubsection{Evaluation Instruction}
Empirical research also guides both the scope and methodology of tool assessments, though it is less commonly observed (7/311).

\stitle{Coverage Validation (6/7).}
Most references aim to align a tool’s evaluation scope with known dimensions from prior studies.
For example, tools like Animated Vega-Lite~\cite{Zong2022}, Data Animator~\cite{Thompson2021}, and AniVis~\cite{Li2021c} demonstrate coverage of the core animation transitions documented by Heer \etal\cite{Heer2007} in their example galleries to showcase their effectiveness.
Data Animator~\cite{Thompson2021} further showcases coverage of the animated data graphic design space~\cite{Thompson2020}.
WonderFlow~\cite{wonderflow} ensures its examples cover all semantic labels in narrations summarized by Cheng \etal\cite{Cheng2022}.
Through this approach, tools emphasize completeness and rigor without necessarily replicating prior experimental setups.

\stitle{Protocol Adaptation (1/7).}
In rare instances, tools integrate empirical evaluation methods directly. 
Visual Foreshadowing~\cite{Li2020c}, for example, adopts subjective questionnaires from Amini \etal\cite{Amini2018a} to systematically measure the effectiveness of its animation techniques. This usage indicates a deeper reliance on empirical procedures but remains unusual.

\textit{Reflection.}
The scarcity of evaluation-instruction citations might reflect the uniqueness of each tool, as authors tailor evaluations to specific tool goals, but it could also suggest missed opportunities for comparability and cumulative learning. 
If two similar tools used common tasks or metrics (with appropriate citations), authors could streamline results and reduce redundant effort.
The current pattern implies an area for growth: the community could benefit from more standardized evaluation practices, enabling future tool authors to reference and build on proven evaluation designs.

\subsubsection{Future Work Discussion}
Empirical research not only shapes current designs but also opens avenues for future exploration (22/311). 

\stitle{Depth Exploration (10/22).}
A significant portion of future work highlights the need for a deeper exploration of dimensions already considered in current tools, though not yet fully explored, such as animation transitions~\cite{Heer2007} for VisConductor~\cite{Femi-Gege2024}, visualization in motion~\cite{Yao2022} for SwimFlow~\cite{Yao2024}, and more diverse and powerful animation effects or principles~\cite{Heer2007,Kim2019c,Thompson2020} for animation grammars~\cite{Kim2020,Kim,Ge2020}. 
Additionally, some works critically discuss the integration of empirical findings. For instance, Datamations~\cite{Pu2021} seeks to further study the effectiveness of animation~\cite{Ferrand2010}, while Lu \etal\cite{Lu2020a} emphasize the trade-offs with animation limitations like attention distraction~\cite{Ferrand2010}. 

\stitle{Horizon Expansion (12/22).}
Another stream of future work calls for the integration of novel aspects not covered in current state, such as cinematic effects~\cite{Xu2023b}, advanced narrative structures~\cite{Lan2021a,Yang2022a}, and emotional effects~\cite{Lan2022}. Many tools~\cite{dataplaywright,wonderflow,dataplayer,datadirector,Li2023b,Shi2021b,Cao2020a} include these new dimensions to further enrich storytelling and user engagement.

\textit{Reflection.}
Observation workshops dominate this function, as they often uncover edge cases and emerging user needs.
When authors reference prior studies in outlining future work, they signal alignment with broader community-identified gaps and view innovation as a continuous process based on established knowledge.

\subsection{Overall Observation}\label{sec:practiceOber}

The structured analysis reveals that empirical research serves both as a foundational framework and a driving force in tool design, shaping decisions from high-level problem formulation to low-level technical implementation. 
We have some key observations as follows:  

\stitle{Diverse Forms and Uses of Empirical Insights.}  
Empirical studies on data videos span a wide array of target components (\eg visuals, motion, narrative, audio) and phenomena (\eg emotion, cinematic effects, character design), and produce findings at varying levels of abstraction—from high-level concepts to low-level technical guidance. Creation tool developers draw on these insights in multiple ways.
Rather than applying studies wholesale, developers selectively adapt and combine findings to fit specific design needs, highlighting the flexible and multifaceted role empirical research plays in tool development.

\stitle{Influence of Empirical Findings: Emphasis on Expressiveness over Learnability.}
Data video creation tools generally aim to balance two goals: \textit{expressiveness} (\ie the range of data videos that can be created) and \textit{learnability} (\ie the ease of use for creators). 
From the citation (or empirical-tool relationship) perspective, we found that most references focus on enhancing expressiveness, such as introducing new components, expanding design spaces or taxonomies, and evaluating example coverage. In contrast, fewer citations contribute directly to improving learnability, such as offering interaction guidance or recognized patterns for automation. 
This imbalance may stem from two factors. First, as noted in Sec.~\ref{sec:corpusOber}, empirical studies focused on learnability are relatively scarce, and expressiveness-centered findings are often harder (or even not appropriate) to adapt into learnability aspects. Second, as discussed in~\cite{DVSurvey}, existing tools prioritize simplifying authoring workflows over expanding expressive capacity, prompting researchers to draw more on prior system designs or implementation-centric studies (\eg for interaction or human-AI collaboration) than on expressiveness-centric data video research.

\stitle{Underutilization of Empirical Findings in Creation Tools.}
The citation analysis reveals a wide range of uses for empirical insights, yet most citations serve to establish context rather than guide functionality or evaluation. Overall, such integration remains underutilized, echoing the finding in a previous study~\cite{DVSurvey}. 
Among the 46 empirical and 48 tool papers, we observed that all tool papers only cite empirical papers 202 times, just 9.15\% of all possible pairs (46$\times$48).
On average, each tool paper cites 4.21 empirical papers (max~=~12, min~=~0). 
Excluding background and future work citations, which do not directly inform tool design, this drops to just 4.03\%, and 1.85 citations per paper (max~=~7, min~=~0).
These relatively low rates, especially in contrast to the total volume of relevant empirical work, suggest that such studies are not yet deeply embedded in tool-research workflows.

Overall, our citation analysis results in a taxonomy detailing how empirical findings are applied in creation tools research, which we hope could provide valuable guidance for future researchers. 
In addition, our analysis indicates a disconnect: abundant empirical knowledge is available, but it is not fully leveraged by creation tools to gain evidence-based guidance. 
Since citation data alone cannot explain the underlying causes, we conducted expert interviews to explore how tool designers engage with empirical studies and what barriers or enablers shape that process. 
The next section presents these insights, offering a human-centered complement to the citation analysis.

\section{Expert Interview}
\label{sec:interview}

Beyond quantitative citation analysis, we further conducted qualitative expert interviews to investigate \textit{Q3: What strategies and factors affect researchers applying empirical research findings in creation tools research processes?} 
While our citation analysis identifies the taxonomy of applying empirical findings, the interviews aim to uncover the underlying reasons, challenges, and missed opportunities. 
Our empirical research landscape and citation analysis results were validated by participants and served as a reference for discussions in the interviews.

\subsection{Participants and Procedure}
We conducted semi-structured interviews with 11 participants (P1-P11, 7 male, 4 female) aged 25–35, including 7 PhD students, 2 researchers in high-tech companies, and 2 faculty members.
All participants are actively engaged in tool-related research and have expertise in data videos, animation, motion graphics, or narrative visualization. 
They have all published papers in top-tier conferences related to visualization, human-computer interaction, or computer graphics (similar to the venues mentioned in Sec.~\ref{sec: corpus}). On average, they have developed 2.67 related research prototypes (SD = 1.61).

Our interview consisted of two main parts. 
First, we introduced the research context and goals to the participants and collected their demographic information and background details on participants' experience with data video creation.
Second, we conducted a retrospective analysis to understand researchers' current practices and challenges. Examples of questions are: \textit{When would you seek empirical research? Whether and how did these findings help address specific challenges? Why (or not) were certain insights ultimately incorporated into the tools?} 
Participants were encouraged to go through every citation function in their tool papers to recall their own experiences, supplementing missed functions if any. 
Whenever possible, the participants were encouraged to share any artifacts that could help reconstruct their creation process, including any materials referenced during their trial-and-error phases, even if they were not included in the final product or publication.
Our empirical research landscape (Tab.~\ref{tab:Corpus}) and citation analysis results (Sec.~\ref{sec:practice}) were also provided as references.
Each interview lasted between 40-60 minutes.

\subsection{Findings}

We conducted a thematic analysis of the interview transcripts using an open coding approach. One author independently coded participants’ responses, and another author cross-checked the codebook, engaging in discussion to ensure consistency.
Based on this coding process, we identified recurring patterns and organized the findings into two core aspects: 
(1) current practices in creation tool research processes, 
(2) key factors that drive or affect the adoption of current practices.
The following subsections report our findings.

\subsubsection{Practices in Creation Tools Research Processes} \label{sec:how}
All participants confirmed that the six citation functions (Sec.~\ref{sec:citationFunc}) reflected when they mentioned empirical research in their papers and captured their decisions during tool development.
Beyond these citation-based strategies, participants described their practical heuristics when utilizing empirical findings, which are seldom explicitly documented in published papers.
We report related findings in two aspects: \textit{When} and \textit{How} empirical findings are applied in creation tools research processes.

\stitle{When to Use Empirical Findings.}
All participants affirmed that empirical research is strategically referenced at key points where evidence can inform design decisions.
Overall, such findings served as \textit{strategic resources} throughout the project for communication and planning:

\utitle{Defining Scope and Priority.} 
We observed that empirical findings helped focus projects on the right problems to solve. 
Early need-finding studies, such as observations and interviews, identified critical user needs, guiding feature prioritization and project scope. 
Similarly, corpus analyses revealed gaps and trends that highlighted design opportunities. Participants stressed the importance of using evidence to frame problems accurately. 
In practice, empirical findings could act as a compass, aligning the team on which user requirements were ``must-haves'' and which could be deferred, and ensuring that the creative effort was directed towards areas of high impact.

\utitle{Early Project Planning.}  
Empirical findings provided insights into necessary components, technical considerations for specific scenarios, and critical user factors (\eg engagement, accuracy, efficiency). 
They could help inform early-stage project planning, including complexity estimation, difficulty assessment, and resource allocation. 
As P7 noted, ``\textit{Data videos are highly complex; introducing new component types in tools can be challenging. Fortunately, empirical research on many components helps me evaluate both novelty and workload.}''
Leveraging such evidence enabled teams to anticipate challenges and clearly define validation goals, resulting in more focused and robust project execution.

\utitle{Deciding Designs.} 
During design and development, empirical insights became useful whenever teams hit a difficult decision point or faced uncertainty, for example, choosing between alternative motion techniques or interface approaches. 
Empirical findings could be used to negotiate and gain buy-in from teammates for design proposals. 
For example, P4 described sharing user study results or literature findings with the team to justify design recommendations, ``\textit{we should do this because research shows X works better,}'' thus lending credibility to the proposal. P5 also noted that ``\textit{Showing the team a peer-reviewed study backs this idea really got everyone on board.}''
These insights served as a neutral reference point, turning subjective opinions into objective discussions. 

\utitle{Evaluation Planning.}
Empirical studies played a role in evaluation planning; as P3 noted, ``\textit{when testing the tool, I sometimes refer to prior user studies for suitable metrics, tasks, and protocols, especially if my study design is similar to your mentioned empirical methods.}'' 
Upon project completion, they revisited implemented features and discussed potential future features explored in empirical research. 

Those strategic uses, closely aligning with our identified citation functions in Sec.~\ref{sec:citationFunc}, underscore the need for accessible and contextually relevant empirical research at each phase.

\stitle{How to Use Empirical Findings.}
Overall, participants engaged in \textit{creative behaviors}, iteratively reinterpreting and adapting empirical findings from multiple sources:

\utitle{Reinterpretation and Adaptation.}
Participants (9/11) frequently reinterpret empirical findings to fit their domain-specific experiences, translating generalized principles into actionable insights for their projects. For instance, a broad guideline such as ``simplify visual information for novice users'' might be reinterpreted specifically as reducing chart density in data videos. Similarly, empirical guidelines were often adapted rather than discarded when initial conditions differed, such as applying data comic-based findings to data video contexts, by tweaking parameters or integrating supplementary knowledge. Such creative reinterpretation and contextual adaptation ensure alignment with practical constraints, user expectations, and their creative vision.

\utitle{Synthesis of Multiple Insights.}
Participants (7/11) often synthesized findings from diverse sources to inspire new ideas. They may not rely on any single study as a silver bullet; instead, they pool evidence from prior papers, user observations in their own tool-specific studies, and personal experience to inform a design decision. 
For example, insights from a corpus analysis of successful videos were merged with user feedback from interviews, resulting in a novel feature that neither source alone suggested. 
Such knowledge assembling and remixing allowed teams to explore broader design spaces and develop solutions that were both evidence-based and innovative.

\utitle{Trial-and-Error Iterative Adaptation.}
Instead of treating research results as fixed rules, participants (6/11) treated them as evolving resources – starting points to be tested, adjusted, or even reimagined as designs progressed. They usually apply a trial-and-error approach.
They implement design recommendations from prior empirical research in projects and then examine how they affect viewers' experiences. Then they can decide whether to refine the designs or replace them with alternatives.
This trial-and-error approach is inherent to creative tool design.
Moreover, the iterative nature of research itself also plays a role, as P1 noted, ``\textit{Sometimes I want to incorporate a highly relevant, empirically studied feature, but integrating it requires substantial additional modeling. Including it in the current paper could divert focus from the core contribution, so I usually position it as future work.}''

\subsubsection{Factors Influencing the Use of Empirical Findings}
\label{sec:why}
Participants highlighted several reasons why they embrace some empirical research findings in their tools. 
Although these reasons naturally vary by project and context, they revealed common factors that influence the application of empirical findings, as elaborated below.

\stitle{Contextual Relevance.}
This factor refers to the extent to which the context of empirical research findings and creation tools research matches.
It is about matches in various aspects, including the problem scenarios, the addressed data video components, narrative targets, intended audience, \etc~
All participants stressed the importance of this factor.
For instance, if an insight (\eg a best practice for visual encoding or an effective narrative pacing technique) aligns with the type of data story they are telling, they find it immediately useful. As P9 explained, ``\textit{If it doesn’t relate to my story, I often don't catch on how to use it right away.}''
Participants noted two frequent difficulties due to contextual mismatches when they attempted to adapt empirical research findings. 
The first is integrating multiple insights within creation tools for a common goal from different empirical research that targets different design choices for different goals (10/11).
Second, empirical studies often target general scenarios, leading to uncertainty about the effects of design recommendations when creation tools are for specific domains such as healthcare, sports, or tourism (5/11).

\stitle{Granularity Level Matching.} 
This factor refers to the alignment between the granularity of empirical findings and design requirements of creation tools in terms of goals, video components, interactions,~\etc~
Empirical studies vary widely in terms of granularity, not only across different papers but even within a single study. 
For example, an interview study might simultaneously yield macro-level phenomena, meso-level user needs, and micro-level design implications. 
Throughout the tool design lifecycle, developers require insights at differing granularities tailored to specific stages and components. 
Ideally, they envision a structured matrix organized by data video components and research granularity for efficient reference. However, the ambiguous, overlapping nature of empirical findings, along with the relatively limited scope of existing research, makes such a matrix difficult to realize and inherently sparse.
Participants, therefore, must manually translate and synthesize insights from various studies and contexts to achieve the required granularity, particularly when computationally actionable technical references are needed\cite{dataplayer, Chen2022c, Li2023b}. This translation process is challenging, time-consuming, and currently lacks systematic support. 
Therefore, empirical insights presented in concrete, actionable formats are valued, such as step-by-step recommendations and computational guidelines that can be directly applied to a data video component.
For instance, Heer \etal\cite{Heer2007} (mentioned by P3, P5, P8) refined high-level animation principles like congruence and apprehension into concrete guidelines, which were widely adopted in follow-up creation tools.
P9 recalled, ``\textit{Sometimes, I come across guidelines that are closely linked to my research. However, directly translating them into code can be quite a challenge. Therefore, additional efforts are usually needed to bridge these high-level guidelines and the tool implementation.}''
Participants expressed a clear expectation for robust frameworks or tools capable of systematically managing, synthesizing, and automatically translating insights into flexible granularity for streamlined knowledge retrieval.

\stitle{Clarity and Accessibility.} 
The way research findings are communicated can greatly affect whether researchers notice and apply them. 
For instance, participants (6/11) noted that if an empirical insight is buried in a long academic paper or conveyed with dense statistics, it might remain overlooked more easily. In contrast, findings shared through intuitive visuals, concise summaries, or ready-to-use toolkits are far more likely to be tried out. For example, a video tool developer might quickly adopt an idea they saw demonstrated in a conference talk or an online tutorial, while overlooking a similar insight hidden in a journal paper. P8 emphasized, ``\textit{When a study comes with a clear demo or visuals, it sticks with you – you see the value immediately.}'' This highlights the importance of making research insights accessible and easy to grasp.
In addition, relevant studies may span multiple disciplines, making them hard to find with simple queries in a literature survey. Domain-specific terminology differences further hinder search, a concept might be studied under a different name in another field. 
Participants expressed a strong desire for enhanced tools or platforms that improve curation, indexing, and cross-disciplinary visibility of empirical knowledge.

\stitle{Conclusiveness and Credibility.} 
Participants tend to adapt empirical findings that are both conclusive and credible. First, if a design recommendation is backed by conclusive evidence, for example, a measured improvement in viewer recall, it carries more weight in design decisions. Second, participants tend to trust empirical findings that are published in reputable venues or aligned with prior experiences. As P10 put it, when an idea is confirmed by both researchers and peers, it feels safer to trust it. Furthermore, if an empirical finding confirms what participants have already discovered through their projects, they’ll readily embrace it as validation. If a design recommendation conflicts with their own experiences, they remain open but cautious, seeking additional validation before making a change. As P4 stated, ``\textit{I lean on research when it backs up what I’ve observed myself; if it clashes with my gut feeling, I need to test it more.}''

\stitle{Feasibility and Constraints.} 
Practical constraints ultimately determine whether an empirical insight can be applied. Even a promising research finding will be ignored if it requires resources or technical capabilities that the research team does not have. 
For instance, an insight calling for complex motion graphics will be set aside if it would slow down production or exceed the tool’s technical capabilities. Participants stressed that an idea must be not only good in theory but also feasible under real-world conditions. Time, budget, skill set, and platform limitations all play a role.

\section{Suggestions and Opportunities} 
Bridging the gap between empirical and tool research requires a joint effort from the entire data video research community. 
Based on our identified key factors that influence the applicability of empirical insights in tool research (Sec.~\ref{sec:why}), we propose community-wide suggestions and specific opportunities accordingly.

At a high level, the entire community would benefit from deeper, ongoing collaborations where researchers can co-develop annotated data video repositories, host co-design workshops, or produce co-authored papers that integrate empirical findings into actionable tool features. 
Such joint initiatives allow researchers to understand real-world constraints while gaining early access to cutting-edge knowledge~\cite{Colusso2019}. 
In parallel, education can support mutual understanding between empirical and tool research. For instance, empirical research might incorporate practitioner-friendly forms of communication (\eg visual summaries or design briefs) while tool developers can benefit from deeper familiarity with empirical research. Cultivating such reciprocal literacy helps build a more cohesive academic ecosystem~\cite{Cao2023}. 
In addition, documenting successful examples of empirical-to-tool translation (\eg curated ``success galleries'') may inspire new pathways and encourage best practices. This paper's framework, taxonomy, and insights can serve as a start.

To enhance \textbf{contextual relevance}, the community can share case studies that demonstrate how insights were transferred across varied scenarios and user groups, highlighting both challenges and strategies for adaptation. Empirical publications, in turn, can provide clearer discussions of generalization limits and domain-specific adjustments. Co-created examples between domain experts and researchers can help detect subtle contextual conflicts early, increasing the likelihood of smooth adoption. AutoDSL tools may also aid in identifying context-appropriate design choices by mapping empirical findings to specific entities, attributes, relations, and rules~\cite{Shi2024b}.

To address \textbf{granularity mismatch}, researchers could publish findings at multiple levels of detail, ranging from broad principles to step-by-step instructions, so a wider audience can select the right granularity. This approach may be further strengthened by co-developing or adopting standardized frameworks (like answer set programming) that encode empirical constraints in machine-friendly formats. 
Leveraging recent advances in generative AI could then automate some of these translations, with researchers verifying outputs for accuracy.

To elevate the \textbf{clarity and accessibility} of empirical findings, researchers are encouraged to provide intuitive explanations, visual demos, or summarized best practices, which would see much broader adoption as indicated in our interviews. 
To improve discoverability, the community could co-develop indexing mechanisms or chatbot-based tools (\eg using retrieval-augmented generation) that clarify terminology and surface relevant insights across disciplines~\cite{Ying2024}. 
Web-based repositories with demos, reference code, and Q\&A sections can further support quick application~\cite{Lan2022, Cheng2022}. 
Moreover, supplementary materials such as short videos or integration examples are also encouraged to illustrate how findings translate into real-world systems.

To improve \textbf{conclusiveness and credibility}, 
we would like to encourage researchers to strengthen the replicability of empirical findings via shared code, open data, or thorough validations and adopt transparent metrics.
Researchers may also reflect on repeated validations or exceptions across different conditions, such as various cultures and study participant groups~\cite{Wu}. 
Meanwhile, tool developers who see divergent results are urged to document these boundary conditions, \eg via open-source demonstrations or pre-print discussions, clarifying which findings generalize under real-world constraints.

To handle \textbf{feasibility and constraint} issues, the core is to communicate the needs for resources and the rationale behind the compromisation clearly.
By openly discussing resource needs and other constraints in empirical publications, researchers can help avoid unrealistic expectations. 
In tool development processes, if certain features are adapted partially due to feasibility issues, researchers should document the decision-making process and the final outcome of the partial adaptation.
It can both benefit future developers who meet similar constraints and encourage the community to address these gaps through continuous improvements. 
Open-source platforms (\eg GitHub) can host incremental attempts, letting researchers refine solutions collaboratively.

\section{Discussions and Future Work}

\stitle{Finer Characteristic of Empirical Findings.}
In Sec.~\ref{sec:empiricalOrg}, we organize empirical research by both empirical methods and data video components. The data video component dimension is relatively fine-grained and aligns with creation tools, whereas the empirical methods dimension remains high-level. We initially aimed to categorize empirical findings by finer-grained attributes, such as granularity or presentation format, but found that these findings were too diverse and detailed for a more elaborate classification, as discussed in Sec.~\ref{sec:corpusOber}. Such an approach would have led to a highly sparse taxonomy, potentially obscuring meaningful clusters in how empirical work supports creation tools. 
As a pioneer study on how empirical data video research influences tool design, we therefore opted for the higher-level perspective of empirical methods. We hope future efforts will refine this approach into more nuanced patterns.
Moreover, organizing findings at multiple granularity levels could support semi-automated adaptation of high-level knowledge into low-level constraints, based on context-specific needs. Other classification angles (\eg research goals, target users, and scenarios) could also be explored to further broaden the ways we structure and understand empirical findings.

\stitle{Scope of Empirical Research.}
Our literature analysis focuses on data video-related research; however, creation tools often draw on a wider range of empirical work, including studies on creation-oriented interactions or human–AI collaboration (Sec.~\ref{sec:corpusOber} and \ref{sec:practiceOber}). In addition, they may incorporate insights from other disciplines, such as psychology, cognitive science, or sociology, depending on particular scenarios and user needs. Although our expert interviews also captured instances from interviewees in which non–data video studies informed tool design, we could not fully document this broader landscape. Extending our framework and taxonomy to cover more expansive corpora, perhaps even larger topics like data storytelling or communicative visualization~\cite{li2025ai}, could strengthen our findings and deepen cross-disciplinary perspectives.

\stitle{Fostering Empirical Finding Generation.}
As discussed in Sec.~\ref{sec:why}, several external factors (\eg contextual relevance and granularity level matching) often limit the direct applicability of empirical findings. Thus, many researchers conducted their formative studies in tool papers to address unmet requirements. 
Investigating how these targeted formative studies compare with empirical efforts could illuminate gaps in current research practices and suggest improvements. 
Our initial observation is that tool builders commonly encounter design problems or user scenarios not yet covered by empirical work, prompting small-scale but highly focused formative investigations. 
For instance, some tool papers perform corpus analyses of sports videos~\cite{Chen2022c} or animated unit visualizations~\cite{DataParticles2023} to inspire new feature sets; others can leverage prior empirical studies on similar topics, such as the narration-animation interplay study~\cite{Cheng2022} for narration-enriched data video tools~\cite{wonderflow,dataplayer}.
Future research can explore patterns across different empirical methods to optimize or automate workflows. 
For example, quantitative and artifact-centric approaches (\eg corpus analysis) often follow recognizable workflows, suggesting human–AI collaboration or pattern-mining tools could streamline analysis and generate guidelines for diverse topics.

\stitle{Generalizability.}
We envision that our research methodology and findings have the potential to benefit broader audiences.
First, researchers can leverage our mixed-method workflow to understand their domain-specific intra-academia gaps.
For example, data storytelling researchers with focuses on other genres can leverage a similar approach as ours with adjusting the underlying component decompositions. 
Second, while our work focuses on intra-academia knowledge transfer, our findings may provide valuable experiences to bridge research-practice gaps.
We note that our identified challenges may become even more pronounced in real-world development, such as less familiarity with research papers and more diverse user needs. 
To address them, categorizing research papers with our taxonomy of operationalizing findings (Sec.~\ref{sec:citation}) might provide resources for practitioners to learn how to transfer research findings into real-world tools.
Future research could explore how these insights scale across scenarios and domains, further refining the interplay between empirical insights and practical tools.

\stitle{Limitations.}
Our study has two limitations. First, citation patterns evolve over time: newer empirical papers tend not to be cited immediately, whereas older work may accumulate citations based on its foundational status. 
This temporal effect could introduce some bias. 
Future work can revisit and update these insights as more papers emerge and accumulate citations.
Second, our paper collection centers on venues like visualization and HCI; relevant empirical studies published in other fields (\eg cognitive science) might not have been fully captured. We did not observe significant omissions, but collecting an interdisciplinary corpus could enrich our findings with cross-disciplinary thinking.

\section{Conclusion}

This paper examined how empirical research on data videos influences creation tools research, factors and strategies shaping this process, and how to improve their integration. Through literature analysis and expert interviews, we mapped the empirical research landscape, developed a taxonomy of integration patterns, identified key practices patterns and influential factors, and proposed actionable suggestions. We hope these insights can inspire future work, fostering stronger synergy between empirical and tool research, and advancing both data video creation and the broader domain of storytelling and authoring tools.


\newpage
\bibliographystyle{abbrv-doi-hyperref}

\bibliography{references}

\begin{thebibliography}{10}

\bibitem{Amini2015}
F.~Amini, N.~H. Riche, B.~Lee, C.~Hurter, and P.~Irani.
\newblock {Understanding Data Videos: Looking at Narrative Visualization through the Cinematography Lens}.
\newblock In {\em Proc. CHI'15}, pp. 1459--1468. ACM, 2015.

\bibitem{Amini2018a}
F.~Amini, N.~H. Riche, B.~Lee, J.~Leboe-McGowan, and P.~Irani.
\newblock {Hooked on Data Videos: Assessing the Effect of Animation and Pictographs on Viewer Engagement}.
\newblock In {\em Proc. AVI'18}, pp. 1--9. ACM, 2018.

\bibitem{Amini2017}
F.~Amini, N.~H. Riche, B.~Lee, A.~Monroy-Hernandez, and P.~Irani.
\newblock {Authoring Data-Driven Videos with DataClips}.
\newblock {\em IEEE Trans. Vis. Comput. Graph.}, 23(1):501--510, 2017.

\bibitem{Archambault2011}
D.~Archambault, H.~Purchase, and B.~Pinaud.
\newblock {Animation, Small Multiples, and the Effect of Mental Map Preservation in Dynamic Graphs}.
\newblock {\em IEEE Trans. Vis. Comput. Graph.}, 17(4):539--552, 2011.

\bibitem{Bradbury2020}
J.~D. Bradbury and R.~E. Guadagno.
\newblock {Documentary Narrative Visualization: Features and Modes of Documentary Film in Narrative Visualization}.
\newblock {\em Inf. Vis.}, 19(4):339--352, 2020.

\bibitem{Brehmer2017}
M.~Brehmer, B.~Lee, B.~Bach, N.~H. Riche, and T.~Munzner.
\newblock {Timelines Revisited: A Design Space and Considerations for Expressive Storytelling}.
\newblock {\em IEEE Trans. Vis. Comput. Graph.}, 23(9):2151--2164, 2017.

\bibitem{Brehmer2020}
M.~Brehmer, B.~Lee, P.~Isenberg, and E.~K. Choe.
\newblock {A Comparative Evaluation of Animation and Small Multiples for Trend Visualization on Mobile Phones}.
\newblock {\em IEEE Trans. Vis. Comput. Graph.}, 26(1):364--374, 2020.

\bibitem{Cao2023}
H.~Cao, Y.~Lu, Y.~Deng, D.~Mcfarland, and M.~S. Bernstein.
\newblock {Breaking Out of the Ivory Tower: A Large-scale Analysis of Patent Citations to HCI Research}.
\newblock In {\em Proc. CHI’23}, pp. 1--24. ACM, 2023.

\bibitem{Cao2020a}
R.~Cao, S.~Dey, A.~Cunningham, J.~Walsh, R.~T. Smith, J.~E. Zucco, and B.~H. Thomas.
\newblock {Examining the Use of Narrative Constructs in Data Videos}.
\newblock {\em Vis. Inform.}, 4(1):8--22, 2020.

\bibitem{DataParticles2023}
Y.~Cao, J.~{L. E}, Z.~Chen, and H.~Xia.
\newblock {DataParticles: Block-based and Language-oriented Authoring of Animated Unit Visualizations}.
\newblock In {\em Proc. CHI '23}, pp. 1--15. ACM, 2023.

\bibitem{Chalbi2020}
A.~Chalbi, J.~Ritchie, D.~Park, J.~Choi, N.~Roussel, N.~Elmqvist, and F.~Chevalier.
\newblock {Common Fate for Animated Transitions in Visualization}.
\newblock {\em IEEE Trans. Vis. Comput. Graph.}, 26(1):386--396, 2020.

\bibitem{Chen2022c}
Z.~Chen, Q.~Yang, X.~Xie, J.~Beyer, H.~Xia, Y.~Wu, and H.~Pfister.
\newblock {Sporthesia: Augmenting Sports Videos Using Natural Language}.
\newblock {\em IEEE Trans. Vis. Comput. Graph.}, 29(1):918 -- 928, 2023.

\bibitem{Chen2022h}
Z.~Chen, S.~Ye, X.~Chu, H.~Xia, H.~Zhang, H.~Qu, and Y.~Wu.
\newblock {Augmenting Sports Videos with VisCommentator}.
\newblock {\em IEEE Trans. Vis. Comput. Graph.}, 28(1):824--834, 2022.

\bibitem{Cheng2022}
H.~Cheng, J.~Wang, Y.~Wang, B.~Lee, H.~Zhang, and D.~Zhang.
\newblock {Investigating the Role and Interplay of Narrations and Animations in Data Videos}.
\newblock {\em Comput. Graph. Forum}, 41(3):527--539, 2022.

\bibitem{Chevalier2014}
F.~Chevalier, P.~Dragicevic, and S.~Franconeri.
\newblock {The Not-so-Staggering Effect of Staggered Animated Transitions on Visual Tracking}.
\newblock {\em IEEE Trans. Vis. Comput. Graph.}, 20(12):2241--2250, 2014.

\bibitem{Chevalier2016}
F.~Chevalier, N.~H. Riche, C.~Plaisant, A.~Chalbi, and C.~Hurter.
\newblock {Animations 25 Years Later: New Roles and Opportunities}.
\newblock In {\em Proc. AVI'16}, pp. 280--287, 2016.

\bibitem{Colusso2019}
L.~Colusso, R.~Jones, S.~A. Munson, and G.~Hsieh.
\newblock {A Translational Science Model for HCI}.
\newblock In {\em Proc. CHI'19}, pp. 1--13. ACM, 2019.

\bibitem{Concannon2020}
S.~Concannon, N.~Rajan, P.~Shah, D.~Smith, M.~Ursu, and J.~Hook.
\newblock {Brooke Leave Home: Designing a Personalized Film to Support Public Engagement with Open Data}.
\newblock In {\em Proc. CHI'20}, pp. 1--14, 2020.

\bibitem{Conlen2023}
M.~Conlen, J.~Heer, H.~Mushkin, and S.~Davidoff.
\newblock {Cinematic Techniques in Narrative Visualization}.
\newblock {\em arXiv}, pp. 1--13, 2023.

\bibitem{Crnovrsanin2021}
T.~Crnovrsanin, Shilpika, S.~Chandrasegaran, and K.~L. Ma.
\newblock {Staged Animation Strategies for Online Dynamic Networks}.
\newblock {\em IEEE Trans. Vis. Comput. Graph.}, 27(2):539--549, 2021.

\bibitem{Dasu2024}
K.~Dasu, Y.~H. Kuo, and K.~L. Ma.
\newblock {Character-Oriented Design for Visual Data Storytelling}.
\newblock {\em IEEE Trans. Vis. Comput. Graph.}, 30(1):98--108, 2024.

\bibitem{Dasu2021}
K.~Dasu, K.~L. Ma, J.~Ma, and J.~Frazier.
\newblock {Sea of Genes: A Reflection on Visualising Metagenomic Data for Museums}.
\newblock {\em IEEE Trans. Vis. Comput. Graph.}, 27(2):935--945, 2021.

\bibitem{Dragicevic2011}
P.~Dragicevic, A.~Bezerianos, W.~Javed, N.~Elmqvist, and J.-D. Fekete.
\newblock {Temporal Distortion for Animated Transitions}.
\newblock In {\em Proc. CHI'11}, pp. 2009--2018. ACM, 2011.

\bibitem{Femi-Gege2024}
T.~Femi-Gege, M.~Brehmer, and J.~Zhao.
\newblock {VisConductor: Affect-Varying Widgets for Animated Data Storytelling in Gesture-Aware Augmented Video Presentation}.
\newblock In {\em Proc. ISS'24}, pp. 1--22, 2024.

\bibitem{Fisher2021}
J.~Fisher, R.~Chang, and E.~Wu.
\newblock {Automatic Y-axis Rescaling in Dynamic Visualizations}.
\newblock In {\em Proc. VIS'21}, pp. 116--120. IEEE, 2021.

\bibitem{Lee}
T.~Ge, B.~Lee, and Y.~Wang.
\newblock {CAST: Authoring Data-Driven Chart Animations}.
\newblock In {\em Proc. CHI'21}, pp. 1--15. ACM, 2021.

\bibitem{Ge2020}
T.~Ge, Y.~Zhao, B.~Lee, D.~Ren, B.~Chen, and Y.~Wang.
\newblock {Canis: A High-Level Language for Data-Driven Chart Animations}.
\newblock {\em Comput. Graph. Forum}, 39(3):607--617, 2020.

\bibitem{Hall2022}
B.~D. Hall, L.~Bartram, and M.~Brehmer.
\newblock {Augmented Chironomia for Presenting Data to Remote Audiences}.
\newblock In {\em Proc. UIST'22}, pp. 1--14. ACM, 2022.

\bibitem{Heer2007}
J.~Heer and G.~G. Robertson.
\newblock {Animated transitions in statistical data graphics}.
\newblock {\em IEEE Trans. Vis. Comput. Graph.}, 13(6):1240--1247, 2007.

\bibitem{Heimerl2016}
F.~Heimerl, Q.~Han, S.~Koch, and T.~Ertl.
\newblock {CiteRivers: Visual Analytics of Citation Patterns}.
\newblock {\em IEEE Trans. Vis. Comput. Graph.}, 22(1):190--199, 2016.

\bibitem{Henry2007}
N.~Henry, H.~Goodell, N.~Elmqvist, and J.~D. Fekete.
\newblock {20 Years of Four HCI Conferences: A Visual Exploration}.
\newblock {\em Int. J. Hum. Comput. Interact.}, 23(3):239--285, 2007.

\bibitem{Herath2023}
A.~Herath, S.~Sallam, Y.~Sakamoto, R.~Gomez, and P.~Irani.
\newblock {Exploring the Design of Social Robot User Interfaces for Presenting Data-Driven Stories}.
\newblock In {\em Proc. MUM'23}, pp. 315--333. ACM, 2023.

\bibitem{Hernandez2016}
M.~Hern{\'{a}}ndez-alvarez and J.~M. Gomez.
\newblock {Survey about Citation Context Analysis: Tasks, Techniques, and Resources}.
\newblock {\em Nat. Lang. Eng.}, 22(3):327--349, 2016.

\bibitem{John2017}
C.~{John W} and C.~{J. David}.
\newblock {\em {Research Design: Qualitative, Quantitative, and Mixed Methods Approaches}}.
\newblock SAGE, 2017.

\bibitem{Kim2023b}
H.~Kim, J.~Kim, Y.~Han, H.~Hong, O.-S. Kwon, Y.-W. Park, N.~Elmqvist, S.~Ko, and B.~C. Kwon.
\newblock {Towards Visualization Thumbnail Designs That Entice Reading Data-Driven Articles}.
\newblock {\em IEEE Trans. Vis. Comput. Graph.}, 30(8):4825--4840, 2024.

\bibitem{Kim2023e}
N.~W. Kim, G.~Myers, J.~Choi, Y.~Cho, C.~Oh, and Y.-S. Kim.
\newblock {Bridging the Divide: Unraveling the Knowledge Gap in Data Visualization Research and Practice}.
\newblock {\em arXiv}, pp. 1--15, 2023.

\bibitem{Kim2019c}
Y.~Kim, M.~Correll, and J.~Heer.
\newblock {Designing Animated Transitions to Convey Aggregate Operations}.
\newblock {\em Comput. Graph. Forum}, 38(3):541--551, 2019.

\bibitem{Kim2020}
Y.~Kim and J.~Heer.
\newblock {Gemini: A Grammar and Recommender System for Animated Transitions in Statistical Graphics}.
\newblock {\em IEEE Trans. Vis. Comput. Graph.}, 27(2):485--494, 2021.

\bibitem{Kim}
Y.~Kim and J.~Heer.
\newblock {Gemini2: Generating Keyframe-Oriented Animated Transitions between Statistical Graphics}.
\newblock In {\em Proc. VIS'21}, pp. 201--205, 2021.

\bibitem{Kong2017a}
H.~K. Kong, Z.~Liu, and K.~Karahalios.
\newblock {Internal and External Visual Cue Preferences for Visualizations in Presentations}.
\newblock {\em Comput. Graph. Forum}, 36(3):515--525, 2017.

\bibitem{Kong2019}
H.-K. Kong, W.~Zhu, Z.~Liu, and K.~Karahalios.
\newblock {Understanding Visual Cues in Visualizations Accompanied by Audio Narrations}.
\newblock In {\em Proc. CHI'19}, pp. 1--13. ACM, 2019.

\bibitem{Kumar2018}
N.~Kumar and N.~Dell.
\newblock {Towards Informed Practice in HCI for Development}.
\newblock {\em Proc. ACM Hum. Comput. Interact.}, 2:1--20, 2018.

\bibitem{Lai2020a}
C.~Lai, Z.~Lin, R.~Jiang, Y.~Han, C.~Liu, and X.~Yuan.
\newblock {Automatic Annotation Synchronizing with Textual Description for Visualization}.
\newblock In {\em Proc. CHI'20}, pp. 1--13. ACM, 2020.

\bibitem{Lan2022}
X.~Lan, Y.~Shi, Y.~Wu, X.~Jiao, and N.~Cao.
\newblock {Kineticharts: Augmenting Affective Expressiveness of Charts in Data Stories with Animation Design}.
\newblock {\em IEEE Trans. Vis. Comput. Graph.}, 28(1):933--943, 2022.

\bibitem{Lan2021a}
X.~Lan, X.~Xu, and N.~Cao.
\newblock {Understanding Narrative Linearity for Telling Expressive Time-Oriented Stories}.
\newblock In {\em Proc. CHI'21}, pp. 1--13. ACM, 2021.

\bibitem{li2025ai}
H.~Li, Y.~Wang, Q.~V. Liao, and H.~Qu.
\newblock {Why is AI not a Panacea for Data Workers? An Interview Study on Human-AI Collaboration in Data Storytelling}.
\newblock {\em IEEE Trans. Vis. Comput. Graph.}, 2025.

\bibitem{Li2021c}
W.~Li, Y.~Wang, H.~Huang, W.~Cui, H.~Zhang, H.~Qu, and D.~Zhang.
\newblock {AniVis: Generating Animated Transitions Between Statistical Charts with a Tree Model}.
\newblock {\em arxiv}, pp. 1--25, 2021.

\bibitem{Li2020c}
W.~Li, Y.~Wang, H.~Zhang, and H.~Qu.
\newblock {Improving Engagement of Animated Visualization with Visual Foreshadowing}.
\newblock In {\em Proc. VIS'20}, pp. 141--145, 2020.

\bibitem{Li2023b}
W.~Li, Z.~Wang, Y.~Wang, D.~Weng, L.~Xie, S.~Chen, H.~Zhang, and H.~Qu.
\newblock {GeoCamera: Telling Stories in Geographic Visualizations with Camera Movements}.
\newblock In {\em Proc. CHI'23}, pp. 1--15. ACM, 2023.

\bibitem{Lin2023a}
T.~Lin, Z.~Chen, J.~Beyer, Y.~Wu, H.~Pfister, and Y.~Yang.
\newblock {The Ball is in Our Court: Conducting Visualization Research with Sports Experts}.
\newblock {\em IEEE Comput. Graph. Appl.}, 43(1):84--90, 2023.

\bibitem{Lu2020a}
M.~Lu, N.~Fish, S.~Wang, J.~Lanir, D.~Cohen-Or, and H.~Huang.
\newblock {Enhancing Static Charts With Data-Driven Animations}.
\newblock {\em IEEE Trans. Vis. Comput. Graph.}, 28(7):2628--2640, 2022.

\bibitem{Mannocci2019}
A.~Mannocci, F.~Osborne, and E.~Motta.
\newblock {The Evolution of IJHCS and CHI: A Quantitative Analysis}.
\newblock {\em Int. J. Hum. Comput. Stud.}, 131(2):23--40, 2019.

\bibitem{Mittenentzwei2023a}
S.~Mittenentzwei, V.~Wei{\ss}, S.~Schreiber, L.~A. Garrison, S.~Bruckner, M.~Pfister, B.~Preim, and M.~Meuschke.
\newblock {Do Disease Stories Need a Hero? Effects of Human Protagonists on a Narrative Visualization about Cerebral Small Vessel Disease}.
\newblock {\em Comput. Graph. Forum}, 42(3):123--135, 2023.

\bibitem{Nam2024}
J.~W. Nam, T.~Isenberg, and D.~F. Keefe.
\newblock {V-Mail: 3D-Enabled Correspondence About Spatial Data on (Almost) All Your Devices}.
\newblock {\em IEEE Trans. Vis. Comput. Graph.}, 30(4):1853--1867, 2024.

\bibitem{Parsons2021}
P.~Parsons.
\newblock {Understanding Data Visualization Design Practice}.
\newblock {\em IEEE Trans. Vis. Comput. Graph.}, 28(1):665--675, 2022.

\bibitem{Pereira2020}
T.~Pereira, J.~Moreira, D.~Mendes, and D.~Goncalves.
\newblock {Evaluating Animated Transitions between Contiguous Visualizations for Streaming Big Data}.
\newblock In {\em Proc. VIS'20}, pp. 161--165, 2020.

\bibitem{Pu2021}
X.~Pu, S.~Kross, J.~M. Hofman, and D.~G. Goldstein.
\newblock {Datamations: Animated Explanations of Data Analysis Pipelines}.
\newblock In {\em Proc. CHI'21}, pp. 1--14. ACM, 2021.

\bibitem{Robertson2008}
G.~Robertson, R.~Fernandez, D.~Fisher, B.~Lee, and J.~Stasko.
\newblock {Effectiveness of animation in trend visualization}.
\newblock {\em IEEE Trans. Vis. Comput. Graph.}, 14(6):1325--1332, 2008.

\bibitem{Rodrigues2024}
N.~Rodrigues, F.~L. Dennig, V.~Brandt, D.~A. Keim, and D.~Weiskopf.
\newblock {Comparative Evaluation of Animated Scatter Plot Transitions}.
\newblock {\em IEEE Trans. Vis. Comput. Graph.}, 30(6):2929--2941, 2024.

\bibitem{DVEva}
J.~Rogers, L.~Shen, A.~Mosca, E.~Peck, M.~Li, A.~Hakone, K.~Potter, and R.~Chang.
\newblock {House Advantage or House of Cards? Stacking the Deck for Data Videos Leads to Null Results}.
\newblock In {\em Proc. CHI EA'25}, pp. 1--14. ACM, 2025.

\bibitem{Rubab2023}
S.~Rubab, L.~Yu, J.~Tang, and Y.~Wu.
\newblock {Exploring Effective Relationships Between Visual-Audio Channels in Data Visualization}.
\newblock {\em J. Vis.}, 26(4):937--956, 2023.

\bibitem{Sajovic2022}
I.~Sajovic and B.~{Boh Podgornik}.
\newblock {Bibliometric Analysis of Visualizations in Computer Graphics: A Study}.
\newblock {\em SAGE Open}, 12(1):1--17, 2022.

\bibitem{Sakamoto2022}
Y.~Sakamoto, S.~Sallam, A.~Salo, J.~Leboe-McGowan, and P.~Irani.
\newblock {Persuasive Data Storytelling with a Data Video during Covid-19 Infodemic: Affective Pathway to Influence the Users' Perception about Contact Tracing Apps in less than 6 Minutes}.
\newblock In {\em Proc. PacificVis'22}, pp. 176--180. IEEE, 2022.

\bibitem{Sallam2022}
S.~Sallam, Y.~Sakamoto, J.~Leboe-McGowan, C.~Latulipe, and P.~Irani.
\newblock {Towards Design Guidelines for Effective Health-Related Data Videos: An Empirical Investigation of Affect, Personality, and Video Content}.
\newblock In {\em Proc. CHI'22}, pp. 1--22. ACM, 2022.

\bibitem{NarrativePlayer}
Z.~Shao, L.~Shen, H.~Li, Y.~Shan, H.~Qu, Y.~Wang, and S.~Chen.
\newblock {Narrative Player: Reviving Data Narratives with Visuals}.
\newblock {\em IEEE Trans. Vis. Comput. Graph.}, pp. 1--15, 2025.

\bibitem{dataplaywright}
L.~Shen, H.~Li, Y.~Wang, T.~Luo, Y.~Luo, and H.~Qu.
\newblock {Data Playwright: Authoring Data Videos With Annotated Narration}.
\newblock {\em IEEE Trans. Vis. Comput. Graph.}, pp. 1--14, 2024.

\bibitem{datadirector}
L.~Shen, H.~Li, Y.~Wang, and H.~Qu.
\newblock {From Data to Story: Towards Automatic Animated Data Video Creation with LLM-Based Multi-Agent Systems}.
\newblock In {\em Proc. GEN4DS‘24}, pp. 20--27. IEEE, 2024.

\bibitem{DVSurvey}
L.~Shen, H.~Li, Y.~Wang, and H.~Qu.
\newblock {Reflecting on Design Paradigms of Animated Data Video Tools}.
\newblock In {\em Proc. CHI'25}, pp. 1--21. ACM, 2025.

\bibitem{dataplayer}
L.~Shen, Y.~Zhang, H.~Zhang, and Y.~Wang.
\newblock {Data Player: Automatic Generation of Data Videos with Narration-Animation Interplay}.
\newblock {\em IEEE Trans. Vis. Comput. Graph.}, 30(1):109--119, 2024.

\bibitem{Shen2024b}
Y.~Shen, H.~Shi, B.~Lee, and Y.~Wang.
\newblock {Authoring Data-Driven Chart Animations}.
\newblock {\em IEEE Trans. Vis. Comput. Graph.}, pp. 1--17, 2024.

\bibitem{Shi2021a}
D.~Shi, F.~Sun, X.~Xu, X.~Lan, D.~Gotz, and N.~Cao.
\newblock {AutoClips: An Automatic Approach to Video Generation from Data Facts}.
\newblock {\em Comput. Graph. Forum}, 40(3):495--505, 2021.

\bibitem{Shi2021b}
Y.~Shi, X.~Lan, J.~Li, Z.~Li, and N.~Cao.
\newblock {Communicating with Motion: A Design Space for Animated Visual Narratives in Data Videos}.
\newblock In {\em Proc. CHI'21}, pp. 1--13. ACM, 2021.

\bibitem{Shi2024b}
Y.-Z. Shi, H.~Li, L.~Ruan, and H.~Qu.
\newblock {Constraint Representation Towards Precise Data-Driven Storytelling}.
\newblock In {\em Proc. GEN4DS'24}, pp. 4--12. IEEE, 2024.

\bibitem{Shin2022}
M.~Shin, J.~Kim, Y.~Han, L.~Xie, M.~Whitelaw, B.~C. Kwon, S.~Ko, and N.~Elmqvist.
\newblock {Roslingifier: Semi-Automated Storytelling for Animated Scatterplots}.
\newblock {\em IEEE Trans. Vis. Comput. Graph.}, 29(6):2980--2995, 2023.

\bibitem{Shu2020}
X.~Shu, A.~Wu, J.~Tang, B.~Bach, Y.~Wu, and H.~Qu.
\newblock {What Makes a Data-GIF Understandable?}
\newblock {\em IEEE Trans. Vis. Comput. Graph.}, 27(2):1492--1502, 2021.

\bibitem{Tang2020}
J.~Tang, L.~Yu, T.~Tang, X.~Shu, L.~Ying, Y.~Zhou, P.~Ren, and Y.~Wu.
\newblock {Narrative Transitions in Data Videos}.
\newblock In {\em Proc. VIS'20}, pp. 151--155. IEEE, 2020.

\bibitem{Tang2020a}
T.~Tang, J.~Tang, J.~Hong, L.~Yu, P.~Ren, and Y.~Wu.
\newblock {Design Guidelines for Augmenting Short-form Videos Using Animated Data Visualizations}.
\newblock {\em J. Vis.}, 23(4):707--720, 2020.

\bibitem{Tang2022}
T.~Tang, J.~Tang, J.~Lai, L.~Ying, Y.~Wu, L.~Yu, and P.~Ren.
\newblock {SmartShots: An Optimization Approach for Generating Videos with Data Visualizations Embedded}.
\newblock {\em ACM Trans. Interact. Intell. Syst.}, 12(1):1--21, 2022.

\bibitem{Thompson2020}
J.~Thompson, Z.~Liu, W.~Li, and J.~Stasko.
\newblock {Understanding the Design Space and Authoring Paradigms for Animated Data Graphics}.
\newblock {\em Comput. Graph. Forum}, 39(3):207--218, 2020.

\bibitem{Thompson2021}
J.~Thompson, Z.~Liu, and J.~Stasko.
\newblock {Data Animator: Authoring Expressive Animated Data Graphics}.
\newblock In {\em Proc. CHI'21}, pp. 1--18. ACM, 2021.

\bibitem{Ferrand2010}
B.~Tversky, J.~B. Morrison, and M.~Betrancourt.
\newblock {Animation: can it facilitate?}
\newblock {\em Int. J. Hum. Comput. Stud.}, 57(4):247--262, 2002.

\bibitem{VanDenBosch2022}
C.~{Van Den Bosch}, N.~Peeters, and S.~Claes.
\newblock {More Weather Tomorrow. Engaging Families with Data through a Personalised Weather Forecast}.
\newblock In {\em Proc. IMX'22}, pp. 1--10, 2022.

\bibitem{Velt2020}
R.~Velt, S.~Benford, and S.~Reeves.
\newblock {Translations and Boundaries in the Gap Between HCI Theory and Design Practice}.
\newblock {\em ACM Trans. Comput.-Hum. Interact.}, 27(4):1--28, 2020.

\bibitem{Wang2021d}
Y.~Wang, Y.~Gao, R.~Huang, W.~Cui, H.~Zhang, and D.~Zhang.
\newblock {Animated Presentation of Static Infographics with InfoMotion}.
\newblock {\em Comput. Graph. Forum}, 40(3):507--518, 2021.

\bibitem{wonderflow}
Y.~Wang, L.~Shen, Z.~You, X.~Shu, B.~Lee, J.~Thompson, H.~Zhang, and D.~Zhang.
\newblock {WonderFlow: Narration-Centric Design of Animated Data Videos}.
\newblock {\em IEEE Trans. Vis. Comput. Graph.}, pp. 1--17, 2024.

\bibitem{Wei2024}
Z.~Wei, H.~Qu, and X.~Xu.
\newblock {Telling Data Stories with the Hero's Journey: Design Guidance for Creating Data Videos}.
\newblock {\em IEEE Trans. Vis. Comput. Graph.}, 31(1):962--972, 2025.

\bibitem{Wu}
A.~Wu, D.~Deng, F.~Cheng, Y.~Wu, S.~Liu, and H.~Qu.
\newblock {In Defence of Visual Analytics Systems: Replies to Critics}.
\newblock {\em IEEE Trans. Vis. Comput. Graph.}, 29(1):1026--1036, 2023.

\bibitem{Xu2023b}
X.~Xu, A.~Wu, L.~Yang, Z.~Wei, R.~Huang, D.~Yip, and H.~Qu.
\newblock {Is It the End? Guidelines for Cinematic Endings in Data Videos}.
\newblock In {\em Proc. CHI'23}, pp. 1--16. ACM, 2023.

\bibitem{Xu2022}
X.~Xu, L.~Yang, D.~Yip, M.~Fan, Z.~Wei, and H.~Qu.
\newblock {From 'Wow' to 'Why': Guidelines for Creating the Opening of a Data Video with Cinematic Styles}.
\newblock In {\em Proc. CHI'22}, pp. 1--20. ACM, 2022.

\bibitem{Yang2023a}
L.~Yang, A.~Wu, W.~Tong, X.~Xu, Z.~Wei, and H.~Qu.
\newblock {Understanding 3D Data Videos: From Screens to Virtual Reality}.
\newblock In {\em Proc. PacificVis'23}, pp. 197--206. IEEE, 2023.

\bibitem{Yang2022a}
L.~Yang, X.~Xu, X.~Y. Lan, Z.~Liu, S.~Guo, Y.~Shi, H.~Qu, and N.~Cao.
\newblock {A Design Space for Applying the Freytag's Pyramid Structure to Data Stories}.
\newblock {\em IEEE Trans. Vis. Comput. Graph.}, 28(1):922--932, 2022.

\bibitem{Yao2022}
L.~Yao, A.~Bezerianos, R.~Vuillemot, and P.~Isenberg.
\newblock {Visualization in Motion: A Research Agenda and Two Evaluations}.
\newblock {\em IEEE Trans. Vis. Comput. Graph.}, 28(10):3546--3562, 2022.

\bibitem{Yao2024}
L.~Yao, R.~Vuillemot, A.~Bezerianos, and P.~Isenberg.
\newblock {Designing for Visualization in Motion: Embedding Visualizations in Swimming Videos}.
\newblock {\em IEEE Trans. Vis. Comput. Graph.}, 30(3):1821--1836, 2024.

\bibitem{Yildirim2023}
N.~Yildirim, M.~Pushkarna, N.~Goyal, M.~Wattenberg, and F.~Vi{\'{e}}gas.
\newblock {Investigating How Practitioners Use Human-AI Guidelines: A Case Study on the People + AI Guidebook}.
\newblock In {\em Proc. CHI'23}, pp. 1--13. ACM, 2023.

\bibitem{Ying2023}
L.~Ying, Y.~Wang, H.~Li, S.~Dou, H.~Zhang, X.~Jiang, H.~Qu, and Y.~Wu.
\newblock {Reviving Static Charts into Live Charts}.
\newblock {\em IEEE Trans. Vis. Comput. Graph.}, pp. 1--16, 2024.

\bibitem{Ying2024}
L.~Ying, A.~Wu, H.~Li, Z.~Deng, J.~Lan, J.~Wu, Y.~Wang, H.~Qu, D.~Deng, and Y.~Wu.
\newblock {VAID: Indexing View Designs in Visual Analytics System}.
\newblock In {\em Proc. CHI'24}, pp. 1--15. ACM, 2024.

\bibitem{Zhao2022}
Z.~Zhao and N.~Elmqvist.
\newblock {DataTV: Streaming Data Videos for Storytelling}.
\newblock {\em arXiv}, pp. 1--16, 2022.

\bibitem{Zong2022}
J.~Zong, J.~Pollock, D.~Wootton, and A.~Satyanarayan.
\newblock {Animated Vega-Lite: Unifying Animation with a Grammar of Interactive Graphics}.
\newblock {\em IEEE Trans. Vis. Comput. Graph.}, 29(1):149--159, 2023.

\end{thebibliography}
\newpage

\end{document}